\documentclass[
reprint,
nofootinbib,
amsmath,
amssymb,
aps, numerical,
longbibliography,
]{revtex4-1}
\usepackage[per-mode=symbol]{siunitx}
\usepackage[dvipsnames,pdftex,hyperref]{xcolor} 
\usepackage[pdftex,unicode,colorlinks,linkcolor=RedViolet,citecolor=BlueViolet,filecolor=magenta,urlcolor=Mahogany]{hyperref}
\usepackage{dblfloatfix}
\usepackage{graphicx}
\graphicspath{{figures/}}
\usepackage{dcolumn}
\usepackage{bm}
\usepackage{hyperref}

\makeatletter
\def\p@subsection{}
\def\p@subsubsection{}
\makeatother

\usepackage{float}
\definecolor{backcolor}{rgb}{0.95,0.95,0.92}
\usepackage{listings}
\usepackage{color}

\definecolor{codegreen}{rgb}{0,0.6,0}
\definecolor{codegray}{rgb}{0.5,0.5,0.5}
\definecolor{codepurple}{rgb}{0.58,0,0.82}
\definecolor{backcolour}{rgb}{0.95,0.95,0.92}
\lstdefinestyle{mystyle}{
	backgroundcolor=\color{backcolor},
	commentstyle=\color{codegreen},
	keywordstyle=\color{magenta},
	numberstyle=\tiny\color{codegray},
	stringstyle=\color{codepurple},
	basicstyle=\footnotesize,
	breakatwhitespace=false,         
	breaklines=true,                 
	captionpos=b,                    
	keepspaces=true,                 
	numbers=left,                    
	numbersep=5pt,                  
	showspaces=false,                
	showstringspaces=false,
	showtabs=false,                  
	tabsize=2
}

\usepackage{chngcntr}
\usepackage{csquotes}
\newcommand{\hyperfootnote}[1][]{\def\ArgI\hyperfootnoteRelay}
\newcommand\hyperfootnoteRelay[2][]{\href{#1#2}{\ArgI}\footnote{\href{#1#2}{#2}}}
\usepackage{afterpage}

\numberwithin{equation}{section}

\usepackage[lofdepth,lotdepth]{subfig}

\begin{document}

\title{A Pedagogic Approach To Visualization In Special Relativity}
\author{Jason Garver}
\email[]{garv0098@umn.edu}
\affiliation{Minnesota Institute for Astrophysics\\
	 University Of Minnesota}
\date{\today}

\begin{abstract}
Special Relativity is often seen as a conceptually difficult topic, which in turn is difficult to effectively teach. This work focuses on the role of visualizations as a tool in teaching Special Relativity at the secondary or university level. The Theory Of Conceptual Fields proposed by G\'erard Vergnaud is used to analyze the conceptualization process in learning Special Relativity so that the ``breakpoints'' in this process can be identified. Visualizations as a didactic tool are then presented to rectify these breakpoints and in general support the conceptualization process. The primary focus here is on Lorentz-FitzGerald contraction in combination with optical distortion due to the finite speed of light, eventually leading to the Penrose-Terrell Rotation effect in three dimensions. Python is used to create a series of images and animations showcasing different objects made of self-luminous rods at various relativistic speeds as seen by an ``observer'', pedagogically so that the breakpoints previously mentioned can be handled. 
\end{abstract}

\begin{titlepage}
\begin{center}

\vspace*{.06\textheight}
{\scshape\Huge \textcolor{Maroon}{University Of Minnesota-Twin Cities}\par}
\vspace{1cm}
\begin{center}
\includegraphics[scale=0.3]{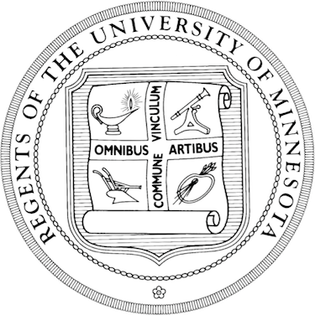}
\end{center}
\vspace{0.4cm} 
\textsc{\huge Astrophysics Thesis}\\[0.5cm]

\hrule \hspace{1pt} \\[0.4cm] 
{\huge \bfseries A Pedagogic Approach To Visualization In Special Relativity \par}\vspace{0.4cm} 
\hrule \hspace{1pt} \\[1.5cm] 

\begin{minipage}[t]{0.4\textwidth}
	\begin{flushleft} \large
		\emph{Author:}\\
		{Jason Garver} 
	\end{flushleft}
\end{minipage}
\begin{minipage}[t]{0.4\textwidth}
	\begin{flushright} \large
		\emph{Supervisor:} \\
		{Dr. Liliya L.R. Williams} 
	\end{flushright}
\end{minipage}\\[3cm]

\vfill

\large \textit{A thesis submitted in fulfillment of the requirements\\ for the degree of Bachelors of Science}\\[3cm]
\vfill
{\large \today}\\[4cm] 
\vfill

\newpage 
\pagenumbering{gobble}
\vspace*{\fill}
\begin{acknowledgments}

This thesis is dedicated first and foremost to Lisa, for supporting my decision to pursue an Astrophysics degree and all the subsequent assistance in editing and proofing my work that ensued.
\newline

I would like to thank Dr. Liliya L.R. Williams for allowing me to explore what amounted to be a very interesting topic in physics education as well as Kaijian Xiao for forcing me to learn Special Relativity properly in the first place.
\end{acknowledgments}
\vspace*{\fill}

\newpage
\end{center}
\end{titlepage}
\maketitle


\tableofcontents

\newpage

\pagenumbering{arabic}
\section{Introduction \label{sec:1}}
In the early 20\textsuperscript{th} century a new proposal was at odds with the prevailing \textit{``Electromagnetic Worldview''}, a theory built around the notion that all physical phenomena can be reduced and explained solely through Maxwell's equations \cite{EMview}. This new idea was, of course, presented in Albert Einstein's paper \textit{``On The Electrodynamics Of Moving Bodies''}\footnote{Historical note: Einstein by no means produced anything completely novel in this paper, rather just an elegant compilation of ideas developed by his contemporaries}, which is now referred to as \textit{``The Special Theory Of Relativity''}. With the help of Lorentz and Minkowski (among others), a new concept of space and time was created on a much more stable footing which described the universe without need for an \ae ther, while more elegantly accounting for important phenomena (ex. the Trouton Experiment(s)\cite{janssen_scirev}) than the former theory stemming from Maxwell's equations could.
\\

In the approximately hundred years since Einstein published the general and special theories of Relativity (and introduced the concepts of the relativity of simultaneity, length contraction, time dilation, etc.) philosophers, scientists and teachers have had to grapple with a new reality which is at its core conceptually different than human experience. These are the conceptual difficulties which arise from seemingly nonsensical ideas about spacetime that make Special Relativity(SR) increasingly difficult to effectively teach. The goal of this work is to give a brief overview an analytic framework for didactics called the Theory of Conceptual Fields \cite{conc}, and how computer generated graphics and animation can be used to effectively illustrate conceptually difficult ideas in (but not limited to) SR. 
\\

Perhaps the most influencing factors in using visualization to understand underlying concepts in SR are what conceptual difficulties students have in the first place. This work will focus on ``relativistic optical distortion'', more simply \textit{what the world would look like} if one were to observe objects moving at relativistic speeds. It is the author's goal to display relativistic phenomena in a simple enough fashion that the core physical ideas are not shrouded by unnecessary visual distractions, and that one conceptual difficulty can be presented at a time. This will be done through a series of animations produced in Python using the power of OpenGL through the package \verb+Vispy+. A wire-frame cube (square, rod, etc) will be observed at a variety of relativistic speeds to explore Lorentz-FitzGerald contraction both with and without optical distortion due to light travel time, as well as investigation of how camera position alters the observed objects (see section \ref{sec:3}). Sample figures can be found in appendix \ref{apdx:B}, and the Python scripts are made availible \href{https://github.com/AstroJasonG/Special-Relativity-Animation}{online here}\footnote{\href{https://github.com/AstroJasonG/Special-Relativity-Animation}{github.com/AstroJasonG/Special-Relativity-Animation}} by the author for personal or professional use. 
\\ 

Finally, detailed sample calculations will be included in appendix \ref{apdx:A} to present \textit{one} method of producing the aforementioned optical distortion that will be appropriate for secondary school teachers, or undergraduates studying physics, as opposed to the formalism of differential geometry and tensors. 

\section{Structure For Conceptualizing Special Relativity \label{sec:2}}
Special Relativity is often regarded as a mathematically dependent and conceptually difficult subject, especially for students inexperienced in physics \cite{difficulties}. This is evidenced most easily in the general lack of the coverage of SR in secondary school settings, and the late introduction of formal SR in the university setting. The author wishes to challenge this idea by pointing out that Newtonian Mechanics, the staple of all secondary and university curricula, can require more mathematical sophistication than SR once one strays from a carefully chosen subset of systems\footnote{i.e. drag, coupled pulleys, double/triple pendulum, etc}. By forming SR using a contextualized approach and more broadly analyzing the didactic sequence using Vergnaud's Theory of Conceptual Fields, these perceived difficulties can be properly accounted for, ultimately allowing more effective instruction in SR at both the secondary and university level.

\subsection{Theory Of Conceptual Fields \label{sec:2.1}}
As an extension of Piaget's theory of human development\cite{piaget}, G\'erard Vergnaud developed the Theory of Conceptual Fields\cite{conc} to describe the learning process of complex and abstract ideas\footnote{Vergnaud focused specifically on children learning mathematics, but the Theory of Conceptual fields extends to any complex learning task}.
\\

Vergnaud's idea was as follows; learning requires interplay between two classes of thinking, what he deems \textit{``situations''}, and by extension \textit{``schemes''} (ways of tackling different situations) and \textit{``concepts''}. As an example of situations and schemes, consider the following which might appear in a math curriculum at the primary level. 

\begin{displayquote}
	Albert is thirty minutes late to dinner with his friend Michele. Michele laments that Albert is always late, and demands an explanation. Albert explains that he had not traveled the usual \SI{1/3}{\kilo \meter} from his home, but instead walked \SI{2}{\kilo \meter} from the patent office where they both work. How much further was Albert forced to walk since he worked late rather than returning home before dinner?
\end{displayquote}

This is an example of a single \textit{``situation''}. Albert is late for supper and the student must determine how much extra walking he had done. Of course one can imagine students solving this particular problem in a myriad of different ways. Perhaps one student counts the number of \SI{1/3}{\kilo \meter} increments between Albert's home, and where he met Michele, when another student simply subtracts the two distances, and yet another counts backwards from \SI{2}{\kilo \meter} to \SI{1/3}{\kilo \meter}. These differing methods are what Vergnaud means by \textit{``schemes''}, or different approaches to a situation. The \textit{``concept''} is of course mathematical manipulation of distances.
\\
\subsubsection{The Conceptual Field \label{sec:2.1.1}}
When one takes a set of situations (which may be only loosely related on the surface) and combines them with a set of concepts, one produces what Vergnaud calls a \textit{``Conceptual Field''}. This relationship between concepts and situations/schemes is bidirectional, as a set of situations cannot all be analyzed with a single concept, and likewise a set of concepts cannot be represented in a single situation. Furthermore, Vergnaud insists that learning requires contrasting situations be analyzed, but also contrasting schemes be used so that the conceptual field is as robust as possible. One simple generalization of the idea of \textit{``schemes''} is symbolic representation\footnote{Which of course is the focus of this work}. In the previous example, a student may choose to draw a pictorial representation of the distances Albert would have traveled in the different situations (leaving from home rather than the patent office). Other students would perhaps express the problem algebraically, using symbols rather than the fractional distances given. In this situation, even the numbers themselves can be represented in various ways, i.e. mixed or improper fractions, or decimal form. These different symbolic representations lead to usage of different schemes in the problem solving process. 
\\

Some important distinctions need to be made about situations, schemes, and concepts before one can progress into the usefulness of the conceptual field. Firstly, schemes must be viewed as general umbrellas to sets of situations. For example, using pictorial representations are not limited to the situation with Albert and Michele, they are a general scheme that can be applied to many situations. Thus a scheme can be defined as having rules and goals, whereas a situation does not have rules or goals initially. Vergnaud also strongly separates concepts and \textit{``theorems''} in the following way. Concepts either apply to a particular situation, or they do not. Theorems however are statements, which can be seen as ``true'' or ``false'' independent of applicability. Concepts are developed through the use of theorems over time, thus a student cannot be ``given'' a concept. Theorems however are able to be handed to the student, and by using theorems appropriately, concepts are built \textit{by the student}.

\subsubsection{Representation And Organization \label{sec:2.1.2}}
The idea of representation is extremely important in science, as the scientist rarely has access to the exact objects of study. In teaching, it is therefore necessary to represent situations and schemes in a way that effectively translates the situation such that a scheme may be applied. Vergnaud outlines two types of representations very important to this work; language/symbols and systems of schemes/sub-schemes.
\\

Language and symbols are the tools of communicating a situation. In the example given previously, the specific language used constitutes one representation of the situation, as well as the form of the information given (the distances). Furthermore, the student may change and manipulate the situation as given using symbols (pictorial, algebraic, graphical). Schemes and sub-schemes\footnote{A sub-scheme simply being a scheme within a scheme} should likewise not be regarded as a static library of information waiting to apply to situations. Instead, when schemes are applied to situations, various sub-schemes can be called upon to further organize incoming information. It is in this behavior, that information obtained in the situation calls on different sub-schemes that makes the organization of information in a situation reflexive and adaptive. As an example of this, one can imagine painting a room. There are a general set of brushes and rollers for which one can use to paint. However, one might discover that painting the corner between two walls is difficult with a roller, and thus switches to a more specialized tool. In this example, the rollers and brushes represent the scheme, the general method to paint the room. The switch to a brush to paint the corners is an example of how new information about the situation can spur reflexive use of a sub-scheme, being the type of brush used.

\subsubsection{Summary \label{sec:2.1.3}}
To conclude the general discussion of the Theory of Conceptual Fields, a more concise definition of concept can be given. A concept is composed of a set of situations, a set of schemes (composed of invariant ideas), and a set of linguistic and symbolic representations. It is proposed by the author that when attempting to conceptualize (build the concept of) something as complex as SR, that it is prudent to analyze these different structures within the concept individually in order to produce an effective didactic sequence that does not leave one area of the concept weaker than others.   

\subsection{Application Of The Theory Of Conceptual Fields \label{sec:2.2}}
Having presented, in an abstract way, the Theory of Conceptual Fields, it will now be prudent to apply the ideas of section \ref{sec:2.1} to SR in a more definite, firm manner. Suppose \textit{concepts} were defined as functions of three variables; the situations $S$, the invariants\footnote{These invariants come in two flavors, concepts-in-action which are important to the particular situation, and theorems-in-action} that make up the schemes $I$, and lastly the representations (linguistic and visual) $R$, then

\begin{align}
	\text{Concept} &= \text{fnc}(S, I, R) \label{eq:2.1}
\end{align}

Having defined \ref{eq:2.1}, the focus of this work can finally take shape and the interplay between visual representations in the concept (and in $S$, $I$) can be explored. The main situations focused on in this work stem from Einstein's  \textit{light postulate}, which states that the speed of light $c$ has a definite and finite value (in a vacuum) of approximately \SI{300000}{\kilo \meter \per \second}. The light postulate is what produces the relativity of simultaneity, and in turn gives rise to the effects of Lorentz-FitzGerald contraction, time dilation, and what will be the main focus here, relativistic optical distortion (Penrose-Terrell Rotation). Thus, a class of situations involving objects moving at relativistic speeds with respect to a stationary \textit{``lab''} observer are of interest in this work.
\\

Furthermore, as the Theory of Conceptual Fields suggests, the ``function'' \ref{eq:2.1} requires situations, invariants, and representation be developed in tandem in order for a high level of conceptualization in the student. It is for this reason that framing visual representations within the conceptual field and the didactic sequence is necessary.\footnote{Context must be given to representations specifically because of the interplay between $S$, $I$, and $R$ mentioned previously} 

\subsection{Role Of Visualization In SR \label{sec:2.3}}
This work is focused primarily on producing clear and informative visualizations to assist in conceptualization of SR, thus the role of visualization must be explored and defined before it can be effectively injected into any didactic sequence. This will require analysis of conceptual difficulties (specifically in regards to the light postulate) students may face with the reality of SR, and how visualizations can dispel them. 

\subsubsection{Conceptual Breakpoints \label{sec:2.3.1}}
As discussed in section \ref{sec:2.1}, holistic conceptualization is built by the use of theorems by means of contrasting schemes and situations over time by the student. To extend this idea, one must realize that physics is largely built upon itself such that no concept a student will receive is completely isolated from another. In more concise language using the \textit{``Arches and Scaffolds''} idea put forth by Janssen\cite{scaffold}, in building SR (an arch), Galilean relativity (and transformations) can be used as a scaffold to support the new constructs (SR). By first understanding non-relativistic kinematics, frames of reference, and Galilean relativity the student can rely upon the scaffold as the transition to the less intuitive Special Relativity is made. Furthermore, introducing Galilean relativity as the scaffold to SR automatically provides situations to compare and contrast. It is for this reason we must seek the ``breakpoints'', places within this process of erecting a scaffold to build the arch (SR), where students are likely to misunderstand or reject new information. Care must be taken at these breakpoints so that the structure does not weaken during construction. Several breakpoints are outlined as follows\cite{strat}.

\begin{itemize}
\item Students have trouble abandoning every-day ``common sense'' when presented with the ``nonsensical'' concepts of SR
\item The $4$-D nature of space-time (as opposed to Euclidean space)
\item Difficulty generalizing the relativistic schemes to the special non-relativistic cases and vice versa
\item Invariance of $c$ and acceptance of its finite value
\item Difficulty connecting new ideas such as invariance and simultaneity to the postulates of SR, and to non-relativistic kinematics
\end{itemize}

The first, second, and fifth points\footnote{This work will only loosely focus on schemes and theorems, thus the third and fourth points are out of its scope} can be directly attacked with visualizations of SR properly contextualized within the didactic sequence. These breakpoints are what this work will attempt to circumvent or strengthen through visualization and animation. Using again the metaphor of \textit{``Arches and Scaffolds''}, it is the aim of this work to not only produce clear pedagogical visualizations, but a series of visualizations that serve as the scaffolding necessary for students to conceptualize SR (keeping in mind the first, second and fifth breakpoints).

\subsubsection{Thought Experiments \label{sec:2.3.2}}
One notable aspect of Albert Einstein's genius was his use of \textit{``thought experiments''} which tend to elude definition, but can be loosely thought of as mental views of situations the experimenter cannot directly observe, but could nonetheless exist in the universe. Scientists are immensely comfortable using thought experiments to form new questions, answer old questions, or critique existing theorems. In essence, thought experiments serve as the playground for the scientist to explore physical reality and ``see what happens''. Special relativity was conceived\footnote{At least to Einstein} on the basis of thought experiments which appear in Einstein's later works on SR and now routinely in SR education.
\\

Thought experiments can play important roles in teaching abstract ideas as are contained in SR\cite{TE}. First and foremost, thought experiments make use of the imagination. The experimenter aims to investigate a situation which is unreachable by every-day experience, with a limited number of theorems to explore a more broad concept, perhaps realizing relationships between situations, or how contrasting schemes could be applied. Unsurprisingly, when students are asked to perform thought experiments, their notions of the situation and what concepts could be important are revealed, allowing the teacher insight into the student's thought process. What's more, thought experiments can be presented to students as a narrative, leading to simple \textit{``if, then''} arguments that can outline the details of a situation clearly. Finally, thought experiments are useful when constructing a scaffold, as abstract situations can be connected to every-day experience\footnote{using language such as \textit{``see''} and \textit{``looks''} to make the situation less abstract} through the narrative of the teacher and the student together.

\subsubsection{Models And Analogies \label{sec:2.3.3}}
A complimentary tool to thought experiments for conceptualizing SR are \textit{``models and analogies''}. Very common in physics, models consist of simpler systems that are easier to understand for the student.\footnote{Example: every system that can be modeled as a simple harmonic oscillator} Analogies in contrast to models, are different situations that once can draw similarities from that are more easily understood. The main hurdle this work faces is the finite value and invariant nature of the speed of light.
When presenting these two ideas, one must rightfully question their validity, origin and consequence within the set of schemes and situations available. To ease this transition, it would be prudent to form an analogy closer to every-day experience. As a simple example of how a ``realistic'' model can be used to help students grasp the finite value of $c$, consider the following.\cite{einsteach}
\\

To acquaint students with the idea of a finite terminal velocity, one can set up a simple demonstration in a real classroom on Earth. Inflate (with air) an ordinary birthday-party balloon and attach a small weight to one end. Then, using either frame-by-frame video analysis, or other techniques, the students can measure the velocity of the balloon as it falls from a set height $h$. Due to the non-trivial surface area of the balloon and speeds involved, air resistance will \textit{not} be negligible and if conditions are all just so, the balloon will approach its terminal velocity. What can be found of course is a relationship between how much gravitational potential energy is expended and the balloon's speed. Little convincing should be required that given a larger height $h$ (and thus more available potential energy), the balloon will asymptotically approach the same maximum velocity as the student can evidence by plotting expended energy versus velocity.
\\

Using the balloon model\footnote{Even ignoring what \textit{causes} air resistance}, the idea of terminal velocity can be extended to relativity as even with an unlimited supply of energy, any spaceship of the imagination will always seem to approach the same terminal velocity $c$. 

\subsubsection{The Train In A Thunderstorm \label{sec:2.3.4}}
Having now defined both analogy and thought experiment in sections \ref{sec:2.3.2} and \ref{sec:2.3.3}, the famous \textit{``Train-Embankment''} thought experiment proposed by Einstein\footnote{And made famous by Landau} can now be posed.

\begin{displayquote}
	Suppose a train travels (parallel) past an embankment with constant velocity $v$ which is a non-trivial fraction of the speed of light. At the precise center of the train sits Michele, and standing on the embankment looking at the train is Albert. Suddenly Albert witnesses two lightning strikes, simultaneously at either end of the train. Michele also experiences the lightning strikes and writes to Albert about the event. When the two patent clerks compare notes, will they agree on the simultaneity of the lightning strikes? 
\end{displayquote}

Einstein gives a rather good explanation of this situation by simply considering the finite value of $c$, if point A is the back of the train, and point B the front:

\begin{displayquote}
\textit{``Now in reality [embankment] he
	[Michele] is hastening towards the beam of light coming
	from B, while he is riding on ahead of the beam of light coming from A. Hence
	the observer will see the beam of light emitted by B earlier than that emitted
	from A. Observers who take the railway train as the reference-body must
	therefore come to the conclusion that the lightning flash at B took place
	earlier than the lightning flash at A. We thus arrive at the important result:
	Events which are simultaneous with reference to the embankment are not
	simultaneous with respect to the train and vice versa.''}\hfill-Albert Einstein\cite{train}
\end{displayquote}

This thought experiment has been used throughout history for good reason, \textit{everybody} knows what a train is. The only new theorem one must accept is the constancy and finite value of $c$, which is rather important when the student is flooded with a lot of contradicting information. However, students and teachers come to dread the train experiment because it's \textit{difficult} to reconcile with every-day experience, and is hard to imagine if not presented very clearly. Throughout the rest of this work, simultaneity will be the main focus in representing situations in SR that have one definite observer (and reference frame) of interest. This situation with a train in a thunderstorm will be extended upon and used as a scaffold\footnote{Perhaps more of a connecting walkway than a scaffold in this case} to other interesting phenomena related to simultaneity.

\subsubsection{Summary \label{sec:2.3.5}}
At the start of section \ref{sec:2.3} the goal was to define the role visualization plays in conceptualizing SR, which required a detour into thought experiments and analogies and more generally what role visualization plays in science. Using careful visual representations of problems and effects will be key for the student to conceptualize the complex nature of SR and space-time. To this end, the author is a proponent of both analogy and thought experiments as representations of SR. However, thought experiments rely on the student's ability to generate situations clearly and correctly. If a student struggles with the first, second and fifth conceptual breakpoints of section \ref{sec:2.3.1}, representation through thought experiment alone will not be sufficient. The author proposes combining thought experiments with real, visual representations of SR by means of computer generated animation. Furthermore, representations commonly used in teaching SR\footnote{Example: Minkowski Diagrams} can either be displayed more simply, interactively, or combined with animation to provide the student with tangible tools to work into schemes. Lastly it will be key to produce visualizations of SR that build upon each other to provide the student with the ``contrasting schemes'' that Vergnaud stresses the importance of in conceptualizing something like SR. 
\\

The \textit{``Arches and Scaffolds''} metaphor presented in section \ref{sec:2.3.1} can be extended again, as visual representations of SR serve as a \textit{``scaffolding arch''}\footnote{Recall the general scaffold from section \ref{sec:2.3.1} was Galilean relativity, but one can imagine using scaffolds in the sense of building a skyscraper where many levels are build on top of another serving as both arches and scaffolds} for conceptualization of SR as a whole. Analogy and metaphor provide the \textit{``springers''}, the stones at either side that form the basis for the arch. Then, simple situations can be represented visually, adding levels of complexity slowly to each visualization to act as the \textit{``voussoirs''}, the stones that make up the bulk of the arch's construction. The student will finally reach the \textit{``keystone''} and the arch will be completed - or in other words, the student will have gained the conceptual background needed to form a large set of varied representations of SR.
\\

This \textit{``scaffolding arch''} makes up part of the $R$ in ``function'' \ref{eq:2.1}, forming the foundation which is then built upon to fully conceptualize SR.

\subsection{Pedagogical Value Of SR And Visualizations \label{sec:2.4}}
Thus far this work has presented a long-winded analysis of how complex concepts like SR are built, and how visual representations apply to that process. This begs the question \textit{``Do students need to conceptualize SR in the first place, and are visual representations, specifically animations, necessary?''}. Special Relativity will admittedly largely not apply to most secondary school student's lives, and only a small subset of university students will find great need for it.
\\

The early 20\textsuperscript{th} century had an explosion of new physics and views about the universe, most notably Quantum Mechanics and Relativity. It is necessary to learn of these new views as part of our \textit{intellectual heritage}, an importance that should be passed onto students. What's more, interest in science can be influenced by the culture and media in which a student has been immersed. Unsurprisingly, students are flooded with headlines of exciting concepts in physics daily from social media. Instead of staying the historical course, education should try to foster this curiosity as early as possible by providing interesting\footnote{and hopefully painless} introductions to ``modern'' physics.\cite{understand}

\subsubsection{A Pedagogical Bridge\label{sec:2.4.1}}
Modern physics is \textit{weird}, quantum mechanics brings a whole host of difficult and crazy ideas to the table that the student \textit{must} drudge through. Newtonian physics in contrast is deeply rooted in what students observe in their world, and as such very little should be truly surprising to the student. Special Relativity has the great ability to be a bridge between these two regimes. Students are acquainted with concepts of motion and velocity in the Newtonian worldview, thus SR can start to acclimate the student to more non-intuitive phenomena. Perhaps most important to the value of teaching SR is the idea that ``common sense'' and every-day experience do \textit{not} always match up with reality - SR can be a ``safe'' way to introduce students to this concept. Furthermore, the historical context behind the development of SR is not only interesting, but also provides an overview of the process in which science is continually changed and revised and insight into the historical and cultural context that surrounded SR.

\subsubsection{The Didactic Sequence \label{sec:2.4.2}}
The role of visualization in a sample didactic sequence\footnote{This sequence is really intended as only a sample. One can imagine another route in which Minkowski space-time is presented before the Lorentz transformations or relativistic kinematics} using the Theory of Conceptual Fields as an analysis tool, can now be explored. Following the work of M. Arlego and M. Otero\cite{didactic}, the sample sequence is as follows:

\begin{enumerate}
	\item Analysis of Galilean reference frames, relative velocity, and Galilean law of addition without a preferential reference frame.
	
	\item Introduction of light postulate and generalize relativity principle. Galilean relativity situations as a scaffold in which to present various schemes that emphasize simultaneity to the student. Then, make the transition to SR by replacing moving objects in the Galilean situations with light beams.
	
	\item Kinematic results of SR, length contraction, time dilation, loss of simultaneity. The Galilean scaffold can now be removed, SR containing all low-velocity limit information.
\end{enumerate}

In order to satisfy the ``concept function'' \ref{eq:2.1}, representations must be present at each of these steps, and more importantly pedagogically linked to situations and schemes as they are presented in the didactic sequence. 
\\

Steps 1. and 2. of the sequence are exactly the ``every-day'' experience students have with relativity, and it is for this reason that computer generated visualizations would \textit{not} be prudent in those cases. Students simply have experience to draw upon, and physical demonstrations can likewise be used. Step 2. does include the transition to SR, but this can be viewed as the scaffolding to the main kinematic results in Step 3.\footnote{Simply because analogy and thought experiment make up the scaffolds for the representations used in SR}
\\

Step 3. is where the bulk of both conceptual breakpoints, and representation difficulties come in. Quite simply, one just does not have a notion of time dilation, length contraction, etc. because we live in a non-relativistic world. The focus on step 3. will include producing animations (in the scaffolding fashion described in \ref{sec:2.3}) with the intent to address conceptual breakpoints one, two, and five before presenting the full relativistic picture. Furthermore, emphasis will be put upon connecting computer generated animation to ``human generated'' animation, i.e. hand drawn diagrams and more importantly visualizations currently used in teaching SR. The intent is for students to use animation and pictorial representation simultaneously in the conceptualization process. This in fact is already a staple of SR learning - \textit{``Now imagine the velocity increases towards the speed of light, how would that change the Minkowski diagram?''}. By providing animations of, for example length contraction, the student can have a tangible experience to draw from when intuition fails.

\section{Animation of SR \label{sec:3}}
The animations of this work were produced using \verb+Vispy+, which combines OpenGL, a very robust and powerful tool for animation, with Python\footnote{While still using some C++ like code}. Python is used opposed to C++ directly because it has become extremely popular as a first programing language. The scripts used here were made to be as widely availible and understandable as possible. \verb+Vispy+ does include a large library of pre-coded objects for animation, however your author wanted the mechanism behind the animation to be extremely transparent, so the shapes animated here are built on a point by point basis ``by hand''. This is done so that if interested, the reader can clearly see how the objects are produced and rendered\footnote{Instead of object abstraction that is seen as more ``Pythonic''} and transformed using SR.
\\

The code of this work comes in three parts, \verb+make_shape.py+, \verb+SR_transforms.py+, and \verb+animate_shape.py+. The script \verb+make_shape.py+ produces a set of points for ``rods'', which are combined to produce a $2D$ square or $3D$ cube.  Next the \verb+SR_transforms.py+ script transforms sets of points by the Lorentz transformations and optical distorion effects (Penrose-Terrell) discussed in \ref{apdx:A2}. Lastly, \verb+animate_shape.py+ pulls all these data together using \verb+Vispy+ to produce interactive visualizations, and animations.\footnote{Vispy can make extremely nice ``interactive'' $3D$ animation, but due to high importance on camera position, that functionality is useless here}

\subsubsection{Goals Of Animation\label{3.0.1}}
At this point it is clear that visual representations are an integral part of the conceptualization process. For this work, a few key areas within SR will be targeted, for example the first, second, and fifth breakpoints of section \ref{sec:2.3.1}. The goal is simple enough, produce animations and graphics which are clear (devoid of unnecessary clutter), interesting and pedagogical.
\\

Firstly this work will include animations of the simplest cases in SR, so that the student may spend more time conceptualizing the physical aspects while acquainting themselves with the mathematical details. Then, to act as the pedagogical bridge mentioned in section \ref{sec:2.4.1}, the animations will focus on showing how non-relativistic scenarios turn into relativistic ones, and vice versa.\footnote{Most importantly showing small $\beta$\footnote{$\beta = v/c$} limits reduce to every day experience}
The simplest independent quantity in SR is the speed of the moving frame. All the interesting phenomena occur at higher speeds, so it is prudent to use animation to cycle through $\beta$ to explore how speed changes how the world appears in relativity.
\\

The major goals of this work are directly related to the conceptual breakpoints of section \ref{sec:2.3.1}. By using animation to show to transition of objects from non-relativistic to relativistic speeds smoothly, the new ``common sense'' (first breakpoint) of SR can be built. With the use of animations of relativistically moving objects paired with more common visuals like Minkowski diagrams, the student can begin to make links between space-time and what physical objects appear to do in space-time (second breakpoint). Lastly, using the calculations behind the animations (appendix \ref{apdx:A}), and using the simplified cases, it can be shown that simultaneity \textit{has} to be a relative concept. Using calculations based on point-by-point analysis it is directly shown how simultaneity needs to be \textit{``forced''} to obtain sensible answers in SR.\footnote{This is one of the main pieces of any problem solving method in SR} Since this stems from the light postulate, using contrasting situations sense can be made of relativistic kinematics.

\subsection{Camera, Observer And Viewing \label{3.1}}
In any visualization of physical phenomena there is a certain degree of \textit{liberty} that must be afforded to it, as if every detail is accounted for the physics would be completely hidden behind other distractions. The intent of these animations are for them to be \textit{simple} so it is prudent at this point to explain the situational details in which liberty is taken. Appendix \ref{apdx:A} gives the mathematical details that correspond to the animations.
\\

The objects discussed in this work can be thought of as constantly self-luminous, that is they are always emitting light in all directions. Effects such as Doppler shifting of the wavelength and aberration are not presented in the animation for the sake of simplicity.\footnote{This however would not be difficult to do with the current form of the animation} For this work it should be assumed that the intensity of light emitted by the object is the same as what is observed, and that the color is ``magically'' corrected so that the observed color is the same as the emitted. Furthermore, considered here are hollow wire-frame objects, the ``wire'' being what is shown in animations.

\subsubsection{Camera And Observer \label{sec:3.1.1}}
Because of the light postulate, the location and nature of the observer plays a huge role in these animations. One must imagine that the ``observer'' does the observing through a camera (as to catch all the very fast details) located at some point $d_{c}$ which will be taken as aligned with the moving frame's $y$-axis (see appendix \ref{apdx:A2}), where the motion is along the $x$-direction. This camera of the imagination can be as sensitive and and fast as required, so that all light that reaches the point $d_{c}$ is captured and recorded. For the general purposes of this work, the camera will be completely stationary (with respect to the lab frame), and will only point ``straight ahead'' along the $y$-axis.

\subsubsection{Viewing \label{sec:3.1.2}}
The next subtlety is how the relativistic objects are \textit{viewed}. In the non-relativistic world, it generally does not make any difference whether a moving object is recorded with the camera following the object, or the camera stationary and the object moving past. However, in SR this is very much not the case. One of the goals of this animation is to show relativistic effects as clearly and simply as possible, which requires the effects be analyzed as independently as possible. It is for this reason that your author has chosen a ``snapshot'' method of viewing, which will be described next.
\\

One can imagine a relativistic box moving past a stationary observer, and since this work focuses on SR and not GR, all frames must remain inertial. Thus, there are two situations available, either the relativistic box is observed at a constant velocity but changing distance with respect to the observer, or the box is observed at a single location in ``snapshots'' at different velocities. To make the latter more tangible, the camera can be thought to be located inside a space ship pointed out a circular window which is only slightly bigger than the appearance of the relativistically moving box. The box moves past the window many times at different constant velocities, and the camera takes one still image each time the box is centered in the window. Then an animation of the box through a variety of speeds can be produced by putting these single images together. This is the situation the following animations describe, as to show how different relativistic speeds change the apparent shape of rods/boxes. The former case where the box's position would change at constant velocity is also interesting, but the effects shown there are non-trivial, and thus left for future work.

\subsubsection{Field Of View \label{sec:3.1.3}}
Another subtlety that has to be taken into account is the camera's field of view. In the $3$D cases, the particular field of view of the camera determines the perspective of the $3$D object, with a narrow field of few corresponding to a telescopic ``lens'', and a large field of view corresponding to a ``macro'' lens. In cinematography and animation field of view is ultimately an artistic decision, but here the choosing of a field of view needs additional thought. 
\\

The main concerns are choosing a ``lens'' for the imaginary camera that firstly does not magnify or de-magnify the object, and secondly yields a visual close to what one would expect with the human eyeball. The ``Angle Of view'' $\varphi$ determines the field of view in the following way.

$$
\varphi = 2 \arctan \frac{l}{2f}
$$

Where $l$ is a characteristic length of the image ``sensor'' and $f$ the focal length of the camera. This relationship is poorly defined in animation, because $f$ is usually found by finding the point in which an image at infinity would be focused for a given lens. One way to set these parameters is as follows\footnote{Which ultimately comes down to personal preference}; let $l = 1$ length units and $f = 1.5$ units (chosen such that the $2f$ point is roughly $3$ times the length of the cube). With these values, $\varphi \approx \SI{36}{\degree}$ which is roughly the same as a typical \SI{35}{\milli \meter} camera. This procedure is very arbitrary, but at least produces a non-magnified image free of noticeable distortion (from the ``lens''). It is worth noting that field of view only becomes important if the object extends into the $\pm y$-directions, and thus it is irrelevant for the $1$D and $2$D cases presented.
 
\subsection{Lorentz Contraction Only \label{sec:3.2}}
In order to present full optical (Penrose-Terrell) distortion, a scaffold of simpler effects must be erected, the simplest of which is the Lorentz contraction. Presented here are simple cases of Lorentz contraction in the ``standard setup'' of appendix \ref{apdx:A}.

\subsubsection{Parallel To Motion \label{sec:3.2.1}}
Consider first a rod emitting light constantly along its length moving parallel to the lab observer's $x$-axis with speed $\beta$. This rod will undergo Lorentz contraction in the $x$-direction only.

\begin{figure}[h]
\centering
	\includegraphics[width=0.4\textwidth]{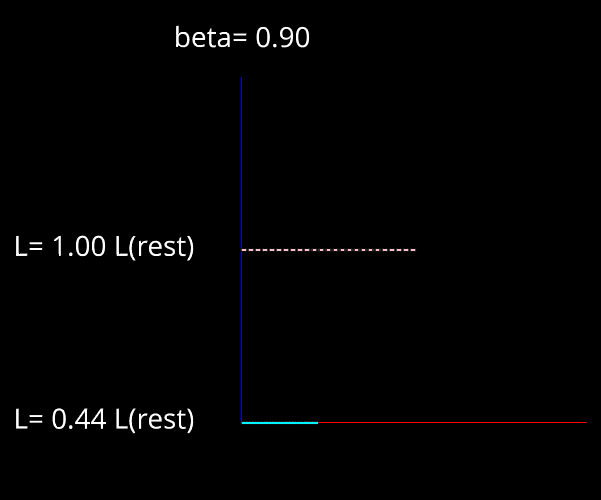}
	\caption{A rod (cyan) oriented parallel to its motion past the lab observer with $\beta = 0.90$ compared to its rest length (dotted pink) due only to Lorentz contraction.}
	\label{fig:1}
\end{figure}

This result is what should be expected by simply dividing the rod's rest length by the Lorentz factor $\gamma$. In this case, simultaneity does \textit{not} need to be considered, as it neglects \textit{how} the observer knows the position of the rod.

\subsubsection{Perpendicular To Motion \label{sec:3.2.2}}
The only other case of Lorentz contraction of rods that must be considered is the case that the rod is perpendicular to the direction of motion. In this case, the rods do not undergo any transformation according to Lorentz. This distinction will become important in the two dimensional case of Penrose-Terrell Rotation.\\

\begin{figure}[h]
	\centering
	\includegraphics[width=0.4\textwidth]{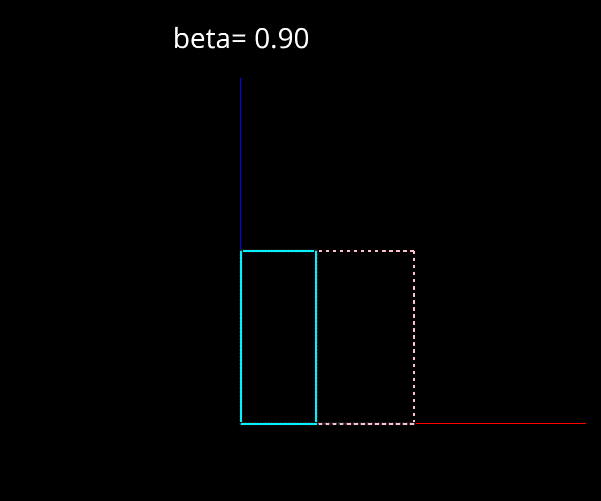}
	\caption{Rods forming a relativistic square ($\beta=0.90$) (cyan), where the parallel rods are parallel to the direction of motion compared to a square at rest (dotted pink)}
	\label{fig:2}
\end{figure}

Figure \ref{fig:2} shows that the perpendicular sides of the square do not change, only the parallel ones as in the previous case.

\subsection{Optical Distortion and Lorentz Contraction \label{sec:3.3}}
When one combines the Lorentz contraction (in the direction of motion) with the fact that the speed of light is constant and finite, one produces the first cases of optical distortion. In these cases all of the Lorentz behavior remains the same, however the ``retarded time''\footnote{See appendix \ref{apdx:A}} produces less trivial effects that must be considered.
\\

At their core, calculations involving retarded time and SR are no different than the situations students face every day in physics; \textit{``If Albert and Michele arrive at the patent office promptly at 8:00 am, and Albert lives further away than Michele, when did Albert leave home?''}. This work is not concerned with two patent clerks in Bern, but rather when and where light must have been emitted from an object to ensure simultaneity of observation.

\subsubsection{Parallel To Motion \label{sec:3.3.1}}
It is again important to start with the most trivial case of optical distortion, a rod oriented parallel to the direction of its motion.\footnote{Still working in the standard situation found in appendix \ref{apdx:A}}. Much like the case of Lorentz contraction alone, this situation is not so difficult to imagine without graphics or animation - the rod's length will contract as a function of its velocity. However, taking into account the travel time of light brings in spatial dependence to the calculation; the rod at constant velocity may change length based on its position relative to the observer! This fact is completely nontrivial, and from whichever scheme one uses to calculate positions of points as seen by the lab frame, it is hard to visualize.
\\

This work will for the moment remain in the standard situation, where snapshots of a moving object are taken directly in front of the lab observer at a variety of speeds $\beta$.

\begin{figure}[h]
	\centering
	\includegraphics[width=0.4\textwidth]{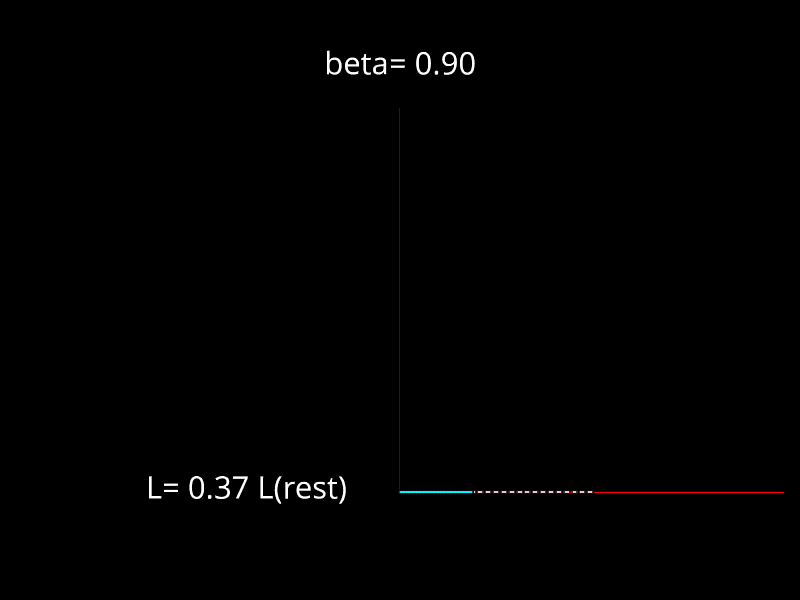}
	\caption{Rod oriented parallel to motion with optical distortion consideration ($\beta=0.90$) (cyan), compared to rod at rest (dotted pink)}
	\label{fig:3}
\end{figure}

The result of comparing figures \ref{fig:1} and \ref{fig:3} may be surprising to the reader, as the optical distortions due to light travel time make the rod \textit{appear} shorter than even its Lorentz contracted length. It should be noted that the rod in figure \ref{fig:3} is already past the observer, which is why it appears shorter; care must be taken in these cases to properly account for the position dependence in the optical distortion.

\subsubsection{Perpendicular To Motion \label{sec:3.3.2}}
The case of rods perpendicular to the direction of motion with optical distortion is the first non-trivial result presented here. Using the standard situation, one must think through \textit{when and where} light must have been emitted by the moving rod to arrive at the observer simultaneously. It should be obvious that since light leaving the ``top'' of a rod oriented perpendicular to the direction of motion travels a further distance to the observer than the bottom. This fact is not so difficult to understand, but given a rod self luminous along its entire length, what shape does the rod appear to be according to the observer?
\\

This difficulty is where the general nature of equation \ref{eq:A3} comes into play. Since \ref{eq:A3} gives the apparent position of \textit{any} point of interest, the behavior of the entire rod can be produced analytically by a simple substitution.\footnote{See appendix \ref{apdx:A} for details} The apparent position of points on a rod perpendicular to the direction of motion evidently form a hyperbola of the form:\footnote{Equation \ref{eq:A4}}

\begin{align}
\frac{\left(x-\gamma A \right)^{2}}{\gamma^{2} \beta^{2} A^{2}}	- 
\frac{\left(y-d_{c}\right)^{2}}{A^{2}} - \frac{z^{2}}{A^{2}} &=1
\end{align}

Where $A = fnc(\beta, \bar{x})$ it is clear these points are points that lie on a hyperbola.

\begin{figure}[h]
	\centering
	\includegraphics[width=0.4\textwidth]{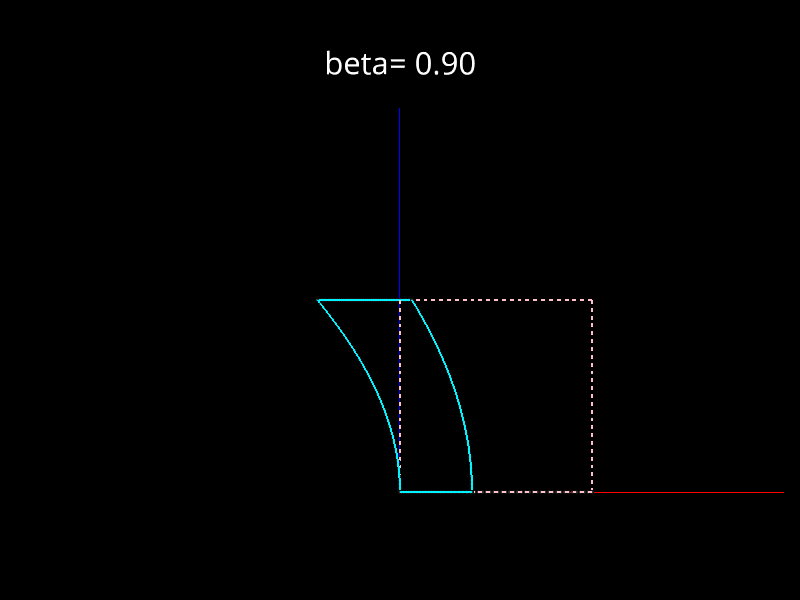}
	\caption{Rods forming relativistic square with Lorentz and optical distortion ($\beta=0.90$) (cyan), compared to square at rest (dotted pink)}
	\label{fig:4}
\end{figure}

Figure \ref{fig:4} shows the perpendicular sides of a square moving along the $x$-axis appear to be bent into the upper branches of a hyperbola.\footnote{Because the bottom edge of the square is aligned with the observer we only see the top branch of the hyperbola}

\subsubsection{Penrose-Terrell Distortion Of 3D Box \label{sec:3.3.3}}
Having shown the effects of optical distortion combined with length contraction for rods parallel and perpendicular to the direction of motion a full 3D box can now be formed.\footnote{In principle any 3D shape can be transformed using equation \ref{eq:A3}} The same principles discussed in the last two examples apply to a 3D box, optical distortions occur because of the finite speed in which light can travel from the cube to the observer.
\\

In 1959, Roger Penrose\cite{penrose_1959} suggested that three dimensional objects moving at relativistic speeds (with the optical considerations) would appear rotated such that a sphere retains its circular profile, but rotates on its axis. Likewise, a cube as is the focus here, would appear to rotate as the \textit{``back''}\footnote{Note: ``back'' is defined as the face with normal pointing in the $-x$-direction, colored cyan. The front being the opposite colored pink} of the cube appears to face the observer. This is an \textit{illusion}, as there is no \textit{``rotation''} in the real sense, only the distortions presented in the previous two examples. This is the author's reasoning in renaming these effects \textit{``distortions''} rather than \textit{``rotations''} as Penrose and Terrell called them. Using the standard setup and transformations as in previous examples, all points on the cube can be visualized. This case is particularly prone to misconception, as the various transformations are applied in 3D there are fewer simplifying symmetries and thus warrants computer aided visualization.
\\

\begin{figure}[h]
	\centering
	\includegraphics[width=0.4\textwidth]{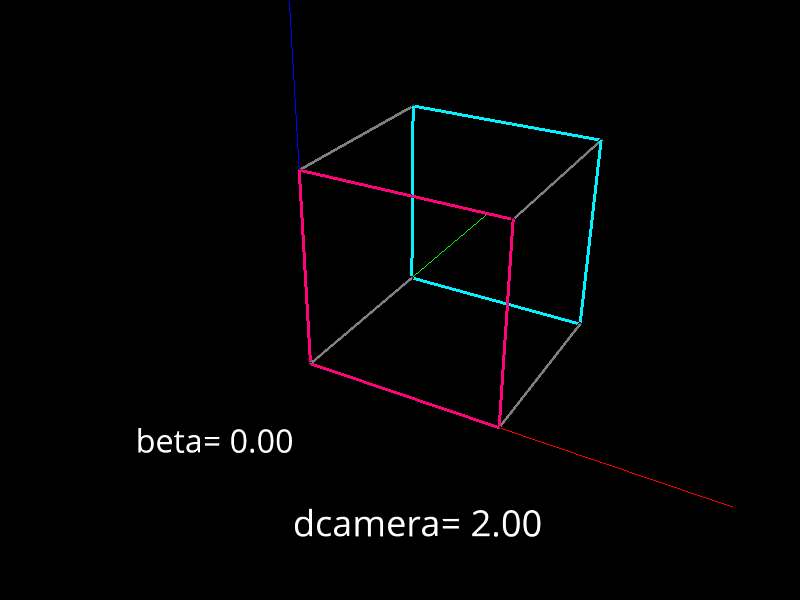}
	\caption{Rotated view of a color-coded 3D cube (at rest) oriented along the $x$-axis. The pink rods will make up what will be called the \textit{``front''} of the cube, while the cyan rods the \textit{``back''}. All gray rods are parallel to the $y$-axis}
	\label{fig:5}
\end{figure}

Figure \ref{fig:5} shows the color scheme of the 3D cube while it is at rest at the origin. However, as seen in equation \ref{eq:A3} the position of the ``camera'' is rather important. In this case, the camera remains at a distance $d_{c}$ along the $y$ axis so it should be noted that the rotated view of figure \ref{fig:5} cannot be used in real visualizations without changing the form of the transformation.\footnote{Thus figure \ref{fig:5} is simply to display the color scheme}

\begin{figure}[h!]
	\centering
	\includegraphics[width=0.4\textwidth]{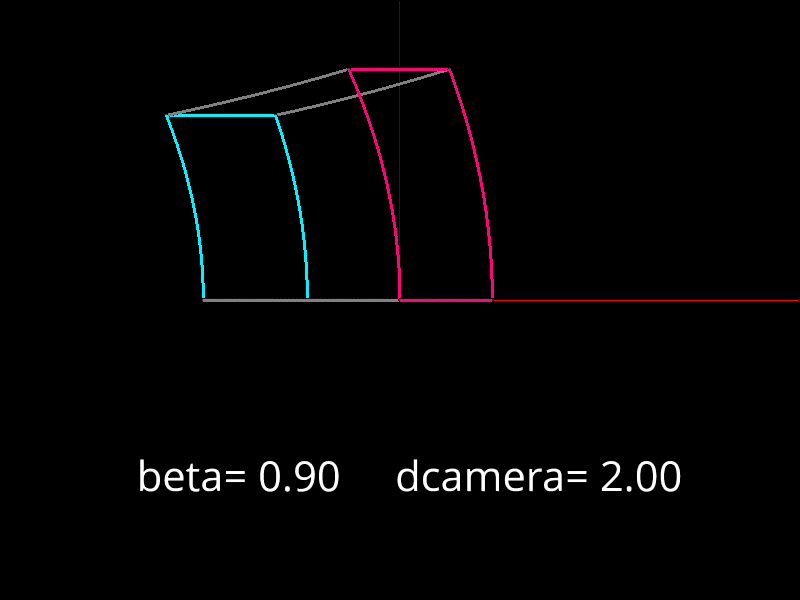}
	\caption{Cube moving with $\beta = 0.90$ and camera distance $2.00$ units along the $y$-axis}
	\label{fig:6}
\end{figure}

As shown in figure \ref{fig:6} and as in the previous examples, the rods of the cube oriented parallel to the direction of motion undergo a \textit{``Lorentz-like''} transformation in that they retain their original shape (of course the length contraction is the same as figure \ref{fig:3}, being somewhat more pronounced than Lorentz contraction). In contrast, the rods perpendicular to motion all appear to be hyperbolas as given in equation \ref{eq:A3}. The apparent ``rotation'' as Penrose was inclined to call it comes from a trivial but less obvious point, the light from the parallel rods forming the far face of the cube are further away than the near face in the $y$-direction \textit{and} further away in the $x$-direction. Since light from more distance parts of the object needs to be emitted earlier to ensure simultaneity, the far face of the cube lags behind the near face. This effect is non-trivial enough that Penrose himself sought to find a rotation-like transformation. The reader should be aware that there is no physical rotation in this scenario, and that this effect is an optical illusion produced by the contraction and optical shift of the far-side rods. 

\begin{figure}[h!]
	\centering
	\includegraphics[width=0.4\textwidth]{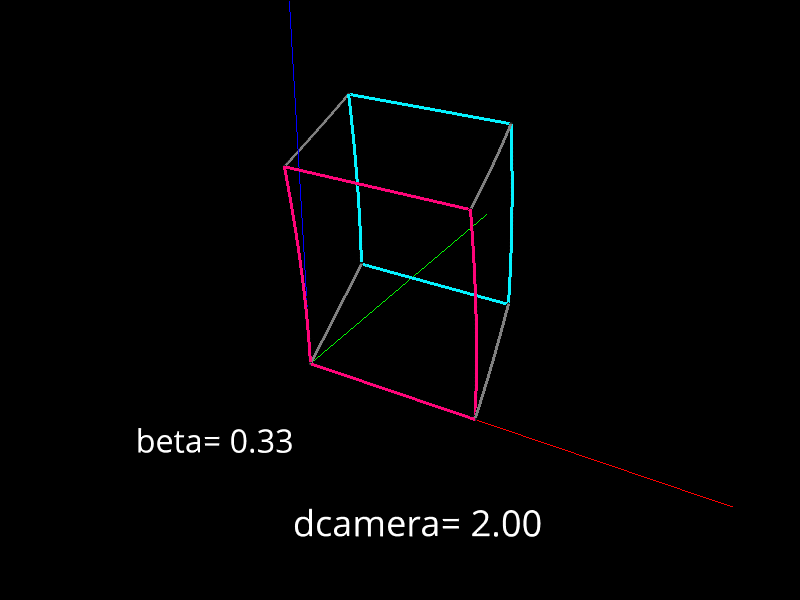}
	\caption{Rotated view of cube moving with $\beta = 0.33$ with respect to the observer. \textit{NOTE:} The coordinates of this cube are the ``extrapolated'' coordinates, that is using the transformations to produce the coordinates mapped to the rotated angle and deprojected to make more clear the ``rotation'' in Penrose-Terrell Rotation is a optical effect, not physical. As stated before, if the camera position were to change, the form of the transformation would also change.}
	\label{fig:7}
\end{figure}

Figure \ref{fig:7} shows the positions of the points of the moving cube are compared to each other (to reiterate, this is not the true representation of a moving cube rotated by some angle, but just computer trickery) the rods parallel to the direction of motion remain parallel, and thus there is no true ``rotation'' as Penrose was seeking.

\subsection{Translations \label{sec:3.4}}
The main focus of this work is the standard situation, where one corner of the object of interest is aligned with the origin at the time of observation. However, in principle the transformation given by \ref{eq:A3} apply to all points, and for any camera position. There are two interesting ways to apply equation \ref{eq:A3} to an object outside of the standard situation, translations side to side, or centered on the origin.
\subsubsection{Centered Object \label{sec:3.4.1}}
Consider a square with the midpoint between all the rods at the origin $(0,0,0)$.

\begin{figure}[h!]
	\centering
	\includegraphics[width=0.4\textwidth]{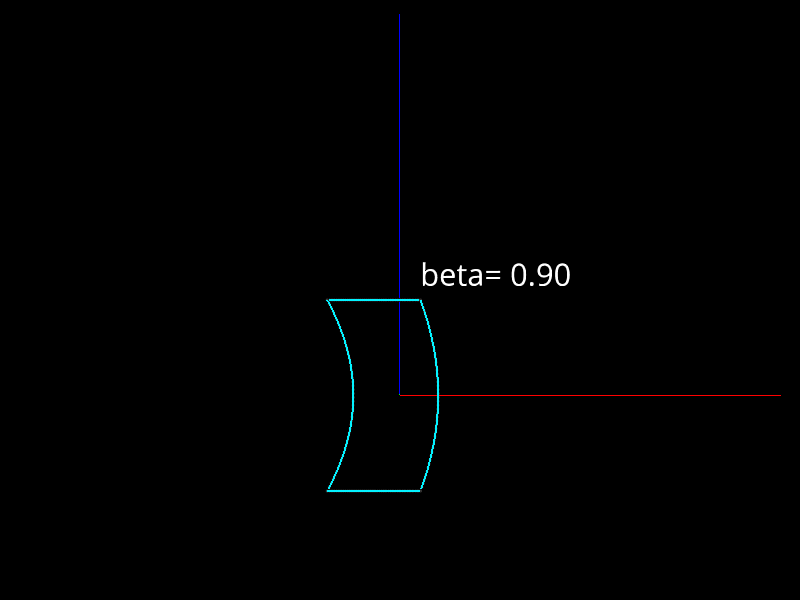}
	\caption{Square centered at the origin moving at $\beta = 0.90$ with respect to the observer}
	\label{fig:8}
\end{figure}

\begin{figure}[h!]
	\centering
	\includegraphics[width=0.4\textwidth]{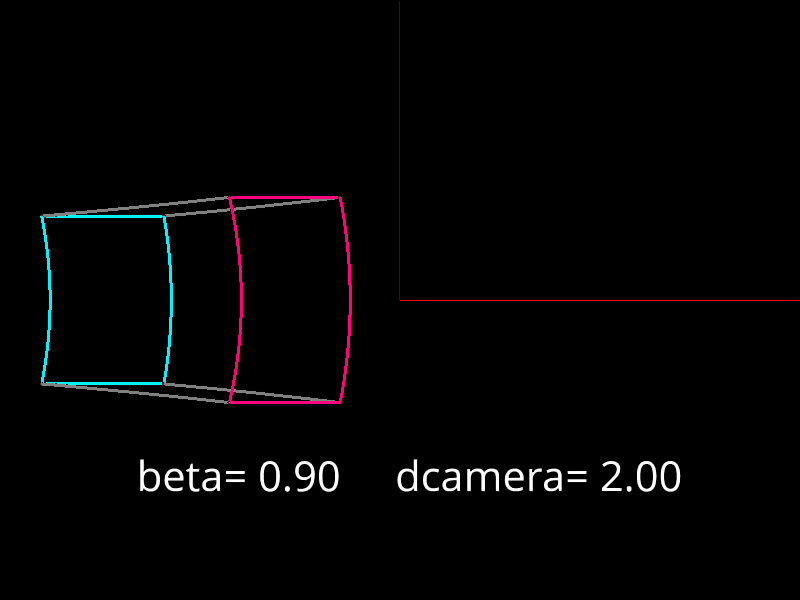}
	\caption{Cube centered at the origin moving at $\beta = 0.90$ with respect to the observer}
	\label{fig:9}
\end{figure}

As seen in figure \ref{fig:8}, the hyperbolic shape given by equation \ref{eq:A4} is symmetric and includes the bottom branch when the object is centered. In the case of a 3D object like figure \ref{fig:9} the transformation becomes less intuitive. Since the cube is centered in all three dimensions, imposing simultaneity causes the cyan rods and the pink rods to be hyperbolas of different eccentricity. This behavior is extremely difficult to visualize accurately without using a computer, and thus is a case where animations of this type excel. 

\subsubsection{Translations Along $x$-Axis \label{sec:3.4.2}}
The second deviation from the standard situation is the object moving past the observer at constant velocity $\beta$. In this scenario, the cube will be observed with its center one unit to the left of the origin, at the origin, and one unit to the right of the origin.

\begin{figure}[h!]
	\centering
	\subfloat[][$\beta = 0.00$]{
		\includegraphics[width=0.23\textwidth]{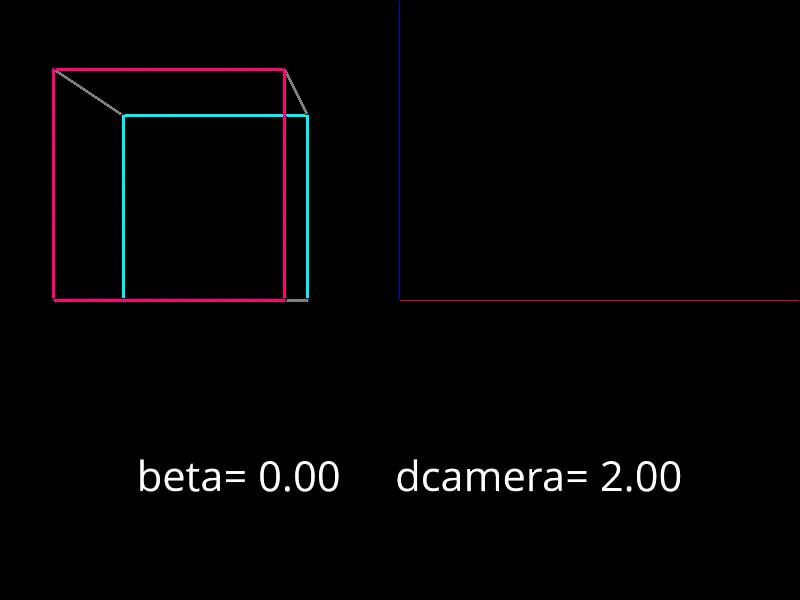}
		\label{fig:10a}}
	\subfloat[][$\beta = 0.33$]{
		\includegraphics[width=0.23\textwidth]{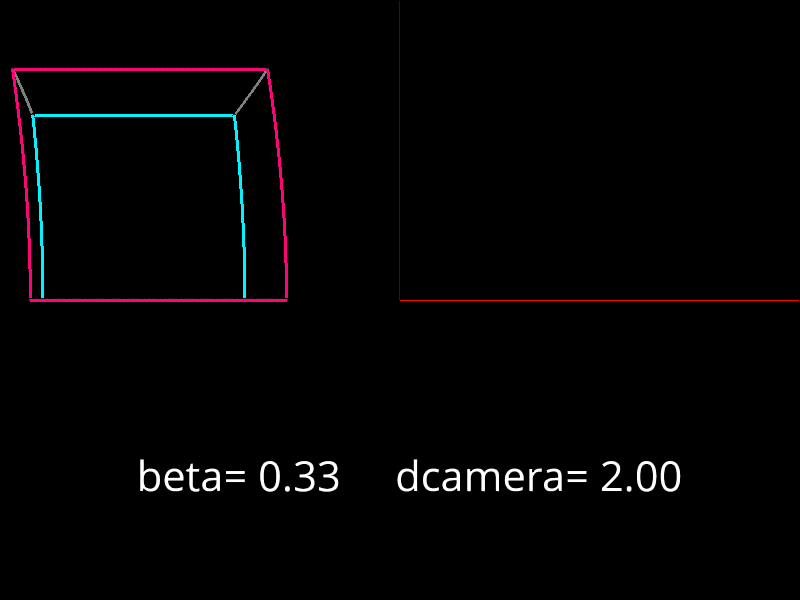}
		\label{fig:10b}}
	\quad
	\subfloat[][$\beta = 0.00$]{
		\includegraphics[width=0.23\textwidth]{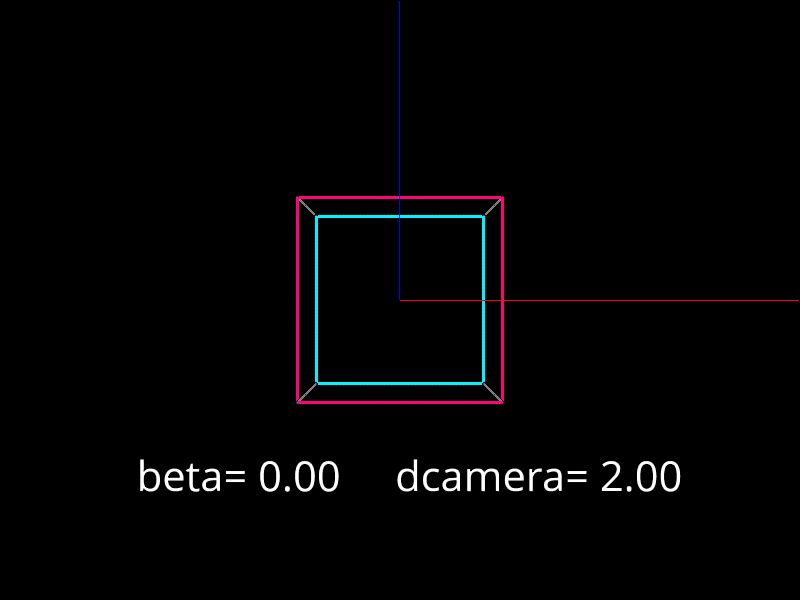}
		\label{fig:10c}}
	\subfloat[][$\beta = 0.33$]{
		\includegraphics[width=0.23\textwidth]{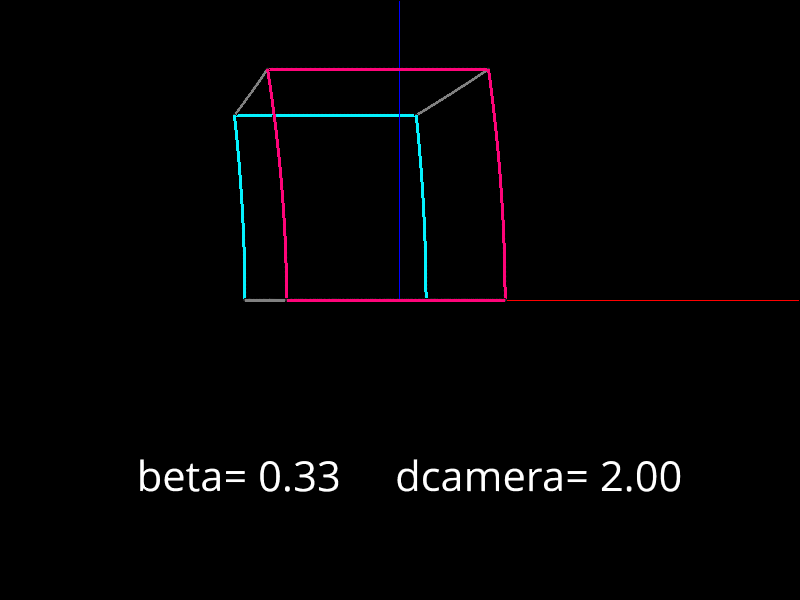}
		\label{fig:10d}}
	\quad
		\subfloat[][$\beta = 0.00$]{
			\includegraphics[width=0.23\textwidth]{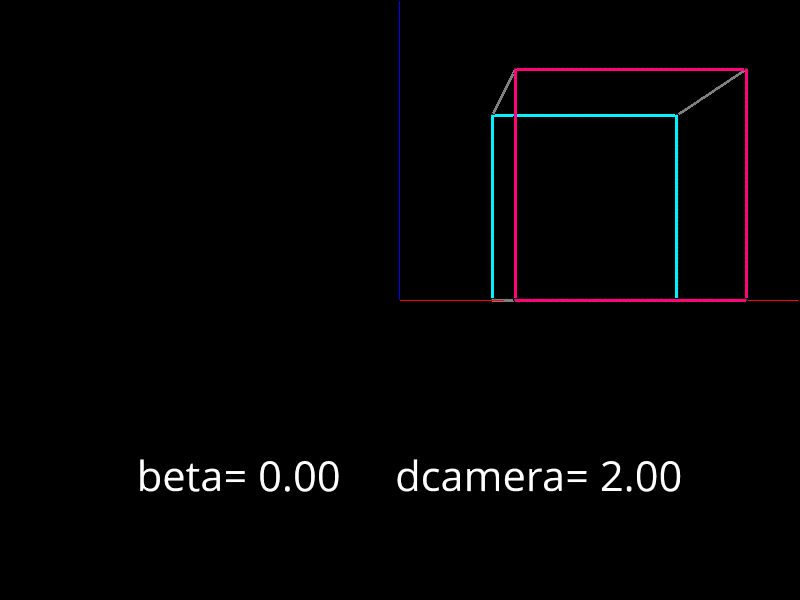}
			\label{fig:10e}}
		\subfloat[][$\beta = 0.33$]{
			\includegraphics[width=0.23\textwidth]{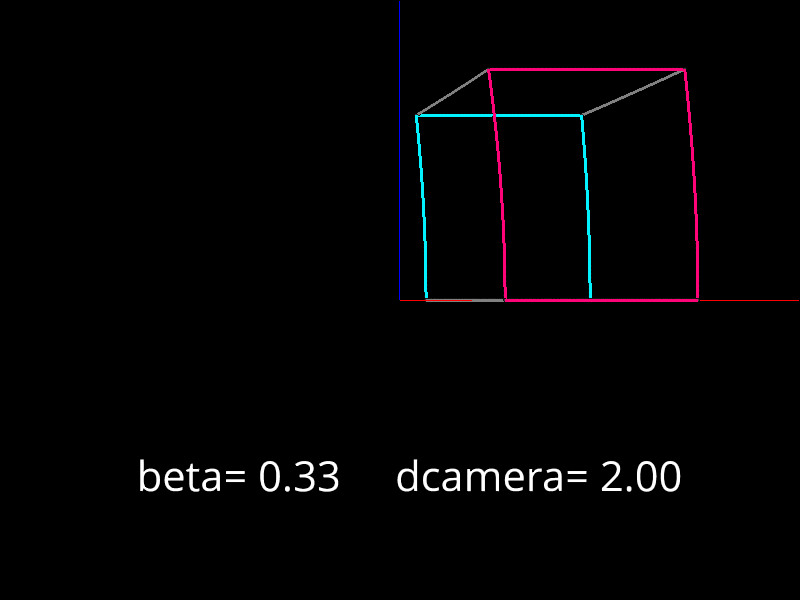}
			\label{fig:10f}}
	\caption{Snapshots of a cube moving past the observer with velocity $\beta = 0.33$ compared to a cube with velocity $\beta \ll 1$ (no relativistic effects)}
\end{figure}

When translations on the $x$-axis are present, the severity of optical distortion of the relativistic cube changes based on position. As seen in figures \ref{fig:10b}, \ref{fig:10d}, \ref{fig:10f}, the optical distortion is less pronounced when the cube is receding from the observer. This can be explained by the fact that the simultaneity condition implies light must leave the cube before the observation time, and in the receding case the position of the cube at the time $t_{r}$ is closer to the origin.\footnote{i.e. light has less distance to travel when it is emitted to the right of the observer} \footnote{In order to ignore forced perspective of the cube, this case must be considered to be very close to the camera such that the horizontal rods remain optically parallel to the $x$-axis. This is not the case if the ``camera'' views a large angle}

\subsubsection{Projection And Moving Camera \label{sec:3.4.3}}
One of the simplifications presented in the standard situation used here is that the observer or ``camera'' does not follow the moving object, but rather takes a snapshot of the object from a fixed position. Furthermore, this work as assumed all the light received by the observer is collected at a single point, which need not be the case. One can define a $u-v$ plane parallel to the $x-z$ plane whose normal may or may not move to follow the center point of the object. The benefit of this method is that a functional form for the projection can be obtained using standard projection methods. Though beyond the scope of this work, the process is explained in detail in reference \cite{reclam}.

\subsection{Discussion \label{sec:3.5}}
The major theme of section \ref{sec:3} was presenting animations which represent one physical effect at a time, as described in the previous section on didactics. Having now presented all of the basic cases of Lorentz Contraction with optical distortion, and what is called the Penrose-Terrell Rotation, these visualizations can be incorporated into the didactics of section \ref{sec:2}.

\subsubsection{A Bridge To New Physics \label{sec:3.5.1}}
Section \ref{sec:2.4.1} describes SR as a safer ``bridge'' to the world of non-Newtonian physics when presented in a clear way. The visualizations of section \ref{sec:3} accomplish this by beginning in the very simplest of cases (one dimensional Lorentz contraction), where only a small leap is required of the student. Then a scaffold of a holistic theory of (Special) Relativity is built using visualizations slowly presenting new physical effects. Not only do the visualizations contribute to the scaffold, but the mathematics presented in appendix \ref{apdx:A} and used in producing the visualizations do as well. Since the calculation giving equation \ref{eq:A3} is general enough to give interesting behavior but simple enough to teach, it can act as additional structure for conceptualizing SR. \footnote{Especially given the point-by-point nature of the mathematics and visualizations, the connection between the two can be stressed more easily}

\subsubsection{Mending Conceptual Breakpoints \label{sec:3.5.2}}
In section \ref{sec:2.3} the main breakpoints in conceptualizing SR are explained, three of which can be actively targeted with animations like the ones in this work. 
\\

The ``first'' breakpoint in section \ref{sec:2.3} describes the initial leap students must take into SR. In every day experience, the properties of space-time are so trivial that they are never outlined in detail. Signals are near-instantly received, space and time are absolute and constant, and one does not have to worry about velocity dependence on the majority of physical effects. Non-relativistic physics are therefore easy to demonstrate in the classroom. SR however is not, and this is the main factor in producing clear visualizations of the concepts like length contraction. Though visualization may not help students philosophically accept new ideas about ``common sense'', connecting mathematical results to more tangible scenarios can begin building that necessary scaffold. This is also where the simplistic nature of visualizations come in, the ``new common sense'' needs to be unobstructed by as many details as possible.
\\

The ``second'' and ``fifth'' breakpoints are handled in a similar way by appropriate visualizations. Somewhere in the didactic sequence for teaching SR, a conversation about simultaneity and space-time has to be brought up. The visualizations of this work specifically use a mathematical framework appropriate for students, so that the same concepts can be derived mathematically \textit{and} shown visually, or vice versa. The point-by-point nature of the calculation in conjunction with the invariance of the speed of light as a postulate allow the student to use the visualization to develop the concept of simultaneity in the following way. Take figure \ref{fig:4} for instance, imposing a constant speed of light \textit{forces} the student to impose simultaneity\footnote{Perhaps on a point-by-point basis} in the same way they would in any kinematic problem like the situation involving Albert and Michele in section \ref{sec:2.1}. This connection to non-Relativistic physics is important to remember, and can be likewise solidified using visualizations that show Relativistic effects from every velocity regime. It is therefore necessary to include limiting behavior in both the simpler Lorentz contraction, and more complicated optical distortion/Penrose-Terrell rotation cases clearly so that the student is reassured that there are no gaps between Newtonian physics, and SR.
\\

One of the most natural extensions from static images to animation is the ability to investigate limiting behavior in \textit{``real time''}. Models, analogies, and thought experiments as explained in section \ref{sec:2.3}, are an integral part of any scientific field, especially physics. First-time physics students, who are not as visually oriented as Einstein, could find undue difficulty in understanding different situations in the form of thought experiments. What visualizations of SR can do, is bring these thought experiments right to the student which can help the student develop a spatial and visual sense of the new physics they're encountering. This skill is particularly important in SR, as the constancy of the speed of light gives a finite range to explore limiting behavior. In the more complex cases like optical distortion and Penrose-Terrell rotation, animation showing a progression from the Newtonian world to the limiting case of $\beta \rightarrow 1$ can highlight subtle physical effects that may have otherwise been hidden behind other complexities.

\subsubsection{Limitations \label{3.5.1}}
As with any physical model, the mathematics and programmatic machinery presented in this work have well defined limitations that must be understood. 
\\

Perhaps the first and most obvious point is the limitation of the mathematics. The derivations of appendix \ref{apdx:A} assume objects are rigid, which is not properly the case in SR due to the finite speed in which information can travel. Thus it has to be noted that the objects studied are either not rigid (held together by some weak imaginary force), or that rigidity is ignored as a simplification. A second limitation of the mathematics is ``real'' rotation. Equation \ref{eq:A3} cannot account for certain scenarios, for example a cube rotating on an axis while having relativistic transverse motion.
Another aspect not included is a moving observer, or both observer and object moving. These effects are less trivial, and not appropriate for the purposes of this work (which is to produce visualizations of the most basic effects in SR).
\\

On the computer science front there is a harsher reality to face, that the simple scripts used in this work do not scale. One difficulty is the production of the relativistic objects. At this time, the author is not aware of stable Pythonic methods to produce any more complex geometric shapes in a vectorized point-by-point format in an efficient way.\footnote{Other geometric figures can be produced, but in a more complex way that hides point-by-point structure} The focus on Python is simple, many students learn Python as a first programming language, and many teachers may also be familiar with it and Object Oriened Programming. A drawback of these methods in Python is if one wishes to include more reference points on the animated objects, the script \verb+animate_shape.py+ runtime will increase linearly with animated points. Furthermore, if the user wishes to produce points in a different coordinate system, for instance spherical, \verb+make_shape.py+ and \verb+SR_transforms.py+ will both require modification to a second order of complexity. The effeciency of the previous two scripts will rapidly decline with more complicated objects or coordinate representations, and thus the package as-is can only be used for relatively simple cases.
\\

Lastly, it should be noted that projects emphasizing visualization in SR and GR (general relativity) are not at all uncommon. The \textit{Open Relativity}\footnote{\href{http://gamelab.mit.edu/research/openrelativity/}{http://gamelab.mit.edu/research/openrelativity/}} project developed at Massachusetts Institute of Technology (MIT)\cite{MITopenrel} is creating an open source first person game, which produces real time manipulation of spacetime based on the player's speed and movement. This framework includes the visuals associated with the Doppler effect, contraction and acceleration in SR/GR. However as one might imagine, putting all of these effects together at once makes for an extremely complicated world view. This was one major contributing factor in the more simplistic nature of this work, while the Open Relativity project does a fantastic job of displaying relativistic effects, its complexity hides the important physics from students new to SR/GR.\footnote{The author would like to note that Open Relativity is great after one has learned SR/GR}

\appendix{}

\section{Mathematics and Derivations \label{apdx:A}}
This appendix is meant as a rough guide to the \textit{schemes} (derivations and mathematical details) used in this work, in particular those which are uncommon or not common at the level an undergraduate or secondary school teacher has access to.
\subsection{Summary of conventions} \label{apdx:A1}
Since there does not exist in Relativity a standard set of notation and convention, included here are what the author uses in this work for the reader's reference.

\begin{description}
	\item[Vectors] It is the author's intent to make this work accessible to lower undergraduate students and secondary school teachers so formulation in terms of tensors will not be used in its entirety. However, it is the author's opinion that even relatively inexperienced students can understand Einstein's Summation convention and should be introduced to it early on. This work will use a hybrid version of tensor notation and standard vector notation.
 	
	\item[Velocity and Lorentz Factor] Where $c$ is the speed of light
		\begin{align*}
			\beta^{i} &= v^{i} / c
			\\
			\gamma &= \frac{1}{\sqrt{1-\beta^{i} \beta_{i}}}
		\end{align*}
	\item[Frames] Any frame which has velocity relative to the ``lab'' frame will be denoted by a ``bar''. e.x.
		\begin{align*}
			\text{lab frame:} \,\,x^{\mu} &\quad \text{moving frame:}\,\, x^{\bar{\mu}} 
		\end{align*}

	\item[Minkowski Metric] 
		\begin{align*}
			\eta^{\mu}_{\nu} &= 
				diag(-1,1,1,1)
		\end{align*}
	
\end{description}
\subsection{Derived Results}\label{apdx:A2}
It is possible categorize computations into three groups termed \textit{``informal''}, \textit{``semi-formal''}, \textit{``formal''} depending on the level of experience necessary to follow them. Loosely, these correspond to the type of calculation typically done in secondary, undergraduate, and post-undergraduate courses. For instance, in SR a formal calculation would require knowledge of tensors and perhaps some differential geometry whereas an informal calculation would amount to algebraically applying Lorentz Transformations. In this appendix I will give results obtained via informal and semi-formal methods, noting that secondary students (and by extension undergraduates) are fully capable of understanding the semi-formal route which while slightly more rigorous, provides more tools to the student.\footnote{As opposed to informal calculations which are generally situation specific and not very useful elsewhere}

\subsubsection{The Standard Setup And Simultaneity}\label{apdx:A2a}
It is paramount in work with SR to very carefully define at what point (in spacetime) measurements are made. This is particularly important because as Einstein's light postulate entails, simultaneity is no longer absolute, and thus we must \textit{force} simultaneity. 
\\

Unless otherwise specifically noted, this work will use the same general setup (a scheme) in all calculations. This is done so that the reader can always reduce a complex situation to a simpler one without any further work, following the discussion of section \ref{sec:2.3.5}. In this standard situation, the observer is placed in the \textit{``Lab Frame''} which we define as stationary with respect to \textit{any} other frame.\footnote{This of course is a forced perspective, in SR there are no special stationary frames} Then, unless noted, the frame that is moving with a constant velocity $\beta^{j}$ with respect to the lab frame will be denoted as the \textit{``Bar Frame''}, e.x. $x^{\bar{\mu}}$. The point of forced simultaneity which will be referenced in every calculation will occur when the origins of the lab and bar frames coincide, which will be taken as time $t = 0$. This point defines a simultaneous event in which to base calculations from.

\subsubsection{Rod Parallel To Motion (Informal):\\ Lorentz With Optical Distortion}
The most simple case of Lorentz contraction with optical distortion (Penrose-Terrell) (displaced perceived position due to finite velocity of light) is the one dimensional rod of rest length $\ell_{0}$ oriented parallel to the $x$-axis of the bar frame with $\beta^{j} = (+\beta, 0, 0)$ with respect to the lab frame (whose $x$-axis is also parallel). What is sought is the length of the rod that an observer at the origin of the lab frame would \textit{see} (Lorentz contraction combined with optical distortion). This calculation will be presented in the \textit{``informal''} fashion, and thus is specific to this geometry and situation.
\\

Firstly let the rod be composed of a ``back'' and a ``front'', define the event when the back end of the rod is precisely aligned with the origin where the observer is located (and the front end is to the right of the origin), at time $t=0$. Suppose the rod always emits light along its entire length and that the observer has the ability to capture this light with a very fast and sensitive camera at the origin of the lab frame. Compared with the known rest length of the rod (in the bar frame), what is the length of the rod that will be observed in the lab frame?
\\

This situation implies that at $t=0=\bar{t}$ the observer has access only to photons currently arriving at the origin, thus the light from the front of the rod was emitted before \textit{now} $t=0$ at time $t_{2} < 0$. Then (at the time the front emitted light) the position of the front of the rod is $x_{2} = ct_{2}$ and the back of the rod $x_{1} = \beta (c t_{2})$, with their sum giving the observed length of the rod. It should be noted that the quantity $\beta(c t_{2})$ is the distance the distance the back of the \textit{rod} has traveled since the front emission whereas $(c t_{2})$ is the distance traveled by the \textit{light} since front emission.

\begin{align}
	x_{1} + x_{2} &= \beta (c t_{2}) + (ct_{2}) \nonumber
	\\
	&= (ct_{2})(\beta + 1) \nonumber
	\\
	&= x_{2}(\beta + 1 ) = \ell_{0} / \gamma \nonumber
	\\ \nonumber 
	\\
	\therefore \qquad x_{1} + x_{2} &= \ell_{0} \bigg[\frac{1-\beta}{1+\beta} \bigg]^{1/2} = \ell \label{eq:A1}
\end{align} 

Where the fact that the rod must undergo Lorentz contraction according the the lab frame is realized. By setting $x_{1} = 0$ (as the problem was posed, the back of the rod is at the origin on observation) the length of the rod $\ell$ in the lab frame is obtained.

\subsubsection{Rod Perpendicular To Motion (Informal):\\Optical Distortion}
The next simplest case is a rod placed perpendicular to the direction of motion. Suppose the situation is the same as the previous example, but now the rod has ``bottom'' initially placed at the origin and ``top'' at a distance equal to the rod's rest length $\ell_{0}$ on the positive $z$-axis, i.e. $x^{\bar{j}} = (0, 0, \ell_{o})$. What does the observer at the origin \textit{see}?
\\

This situation is different than the first in two important ways; the rod does not undergo Lorentz contraction, and the position of points on the rod will now extend to the $x$-$z$ plane. The scheme used here is however essentially the same, if the bottom of the rod is observed at the origin at $t=0$ (meaning the emitted light from the bottom was instantly received by the observer, the forced simultaneous event) then light emitted from the rest of the rod must have left sooner. Suppose the top of the rod emits light at $t_{1} <0$ at point $(x_{1} , 0, \ell_{0})$. It suffices to find the distance of light travel as in the first example. Since Minskowski spacetime is flat, computing this distance is as follows

\begin{align}
	(x_{1})^{2} + (\ell_{0})^{2} &= (c t_{1})^{2} \nonumber
	\\
	&=\left(c\frac{x_{1}}{c\beta}\right)^{2} \nonumber
	\\\nonumber
	\\
	\therefore \qquad x_{1}^{2}&= \frac{(\beta \ell_{0})^{2}}{1-\beta^{2}} \,\, \Rightarrow \,\, x_{1} = \gamma \beta \ell_{0} \label{A2}
\end{align}

Since $t_{1} < 0$, the apparent position of the top of the rod according to an observer at the lab frame origin ($t=0$) is $(\mp \gamma \beta \ell_{0}, 0, \ell_{0})$.

\subsubsection{Optical Distortion And Penrose-Terrell Rotation (Semi-Formal)\label{apdx:A2d}}
In the previous two examples, extremely simple calculations showed that rods parallel and perpendicular to the direction of motion \textit{appear} to have positions different from what one might expect given Lorentz contraction alone. However simple these calculations may be, they are limited. Firstly, the rods were placed in trivial positions, so the problems were at worst two dimensional. Secondly, the observer's position was ``on top of'' the rod's end at the origin, something that is not very realistic. To extend these effects, one should consider the observer or ``camera'' some distance $d_{c}$\footnote{For simplicity the lab axes can always be aligned with the camera along one direction} from the origin, as well as any general point from which light is emitted. Using the scheme from the previous two examples, it is not obvious how to do this in general.
\\

This problem was first solved by Penrose\cite{penrose_1959} and Terrell\cite{terrell}, where it was shown the ``photograph'' of a relativistically moving sphere will retain its circular profile and that objects \textit{appear} to undergo rotation at relativistic speeds, respectively. This effect dubbed \textit{``Penrose-Terrell Rotation''} is subject to great confusion even within the literature. Suppose there is a three dimensional object moving with a constant velocity with respect to an observer. As shown in the first two examples of this appendix, to ensure simultaneity, some observed light must have been emitted at an earlier time, and thus objects appear to be distorted. This effect is a combination of Lorentz contraction and the optical ``lag'' of the light traveling to the observer, and only \textit{appears} to be a rotation in three dimensions. Thus your author would like to rename this effect \textit{``Penrose-Terrell Distortion''} as to avoid the confusion.
\\

To produce the apparent positions of arbitrary points in $3D$, a more rigorous approach is warranted. While there are many methods available (see \cite{apparent,geometry,geom,removal}) this work will give one \textit{``semi-formal''} scheme based on the concept of ``retarded time'' which an undergraduate student might be accustomed to from unrelated calculations involving moving charged particles and their fields.
\\

For this calculation let the observer (camera) exist at (with respect to the origin) $d^{i} = (0, d_{c}, 0)^{T}$, while some arbitrary point of interest is at $P^{i} = (x, y, z)^{T}$. The retarded time can be defined in the following way $t_{r} = (r^{i}\eta_{i}^{j}r_{j})/c$ where $r^{i}\eta_{i}^{j}r_{j}$ is the distance between the camera and point of interest.\footnote{This of course is just the inner product using index notation and the Minkowski metric} In terms of the lab frame's coordinates, the retarded time is thus

\begin{align}
	t_{r} &= -\frac{1}{c} \Big[(x^{2} + (y-d_{c})^{2} + z^{2})^{1/2} - d_{c}\Big] \nonumber
\end{align}

Now, to obtain an expression for $\bar{t}_{r}$, it must be realized that for one, the retarded time will undergo time dilation, and secondly the coordinates that it depends on must be Lorentz transformed, this process is presented next.\footnote{note $d_{c} = ct$}

\begin{align}
	\Lambda^{\bar{\mu}}_{\nu}x^{\nu} &= 
	\begin{pmatrix}
		\gamma& -\gamma \beta_{x}& 0& 0\\
		-\gamma \beta_{x}& \gamma& 0& 0\\
		0& 0& 1& 0\\
		0& 0& 0& 1
	\end{pmatrix}
	\begin{pmatrix}
		ct\\
		x\\
		y\\
		z
	\end{pmatrix} \nonumber
	\\
	\bar{x} &= \gamma (x - \beta c t) \,\, \Rightarrow \,\, \gamma x = \bar{x} + \gamma \beta d_{c} \nonumber
	\\
	\gamma t_{r}&= -\frac{1}{c} \Big[\left([\bar{x}+\gamma \beta d_{c}]^{2} + [\bar{y}-d_{c}]^{2} + \bar{z}^{2}\right)^{1/2} - \gamma d_{c}\Big] =\bar{t}_{r} \nonumber
\end{align}
Using again the Lorentz Transformation, the apparent $x$ position in the lab frame is obtained.

\begin{align}
	x = \gamma (\bar{x} - \beta c \bar{t}_{r}) 
	&= 
	\gamma \Big[(\bar{x} + \beta \gamma d_{c}) - \beta \left[(\bar{x}^{2} + (\bar{y} - d_{c})^{2} + \bar{z}^{2})\right]^{1/2}\Big] \label{eq:A3}
\end{align}

Equation \ref{eq:A3} takes into account both Lorentz and optical distortion effects for any general point.\footnote{recall that the $y$ and $z$ coordinates do not undergo any Lorentz transformation} It should be noted that, this calculation is more involved than the informal cases presented previously, however it is also much more general and uses a scheme that is applied to situations elsewhere in physics.\footnote{Thus either solidifying a student's prior knowledge, or introducing a useful method} With \ref{eq:A3}, a student can find the apparent position of \textit{any} (as long as the object can be presented by a set of points) collection of points in the lab frame and thus are not limited to poles and barns. This result can also reduced to the two informal examples presented, which can be used as a computational check.
\\

As a last note, the behavior of \ref{eq:A3} can be extracted if one sets $A = \bar{x} + \beta \gamma d_{c}$ \cite{scottviner}, then

\begin{align}
\frac{\left(x-\gamma A \right)^{2}}{\gamma^{2} \beta^{2} A^{2}}	- 
\frac{\left(y-d_{c}\right)^{2}}{A^{2}} - \frac{z^{2}}{A^{2}} &=1 \label{eq:A4}
\end{align}

Where for notation sake $\bar{y} = y$, $\bar{z} = z$ since these coordinates do not undergo Lorentz contraction. Equation \ref{eq:A4} is in the form of a hyperbola, thus any line perpendicular to the motion will appear hyperbolic.
\section{Additional Figures \label{apdx:B}}

(See next page)
\newpage
\clearpage
\counterwithin{figure}{section}
\setcounter{figure}{0}

\begin{figure}[H]
\centering
\subfloat[Subfigure 1 list of figures text][ $\beta = 0.00$ ]{
	\includegraphics[width=0.24\textwidth]{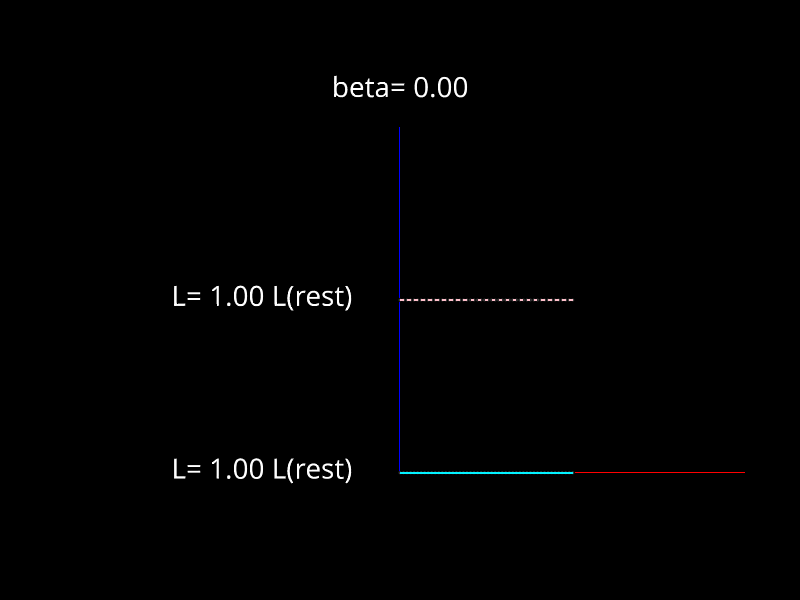}
	\label{fig:B1a}}
\subfloat[Subfigure 2 list of figures text][$\beta = 0.33$]{
	\includegraphics[width=0.24\textwidth]{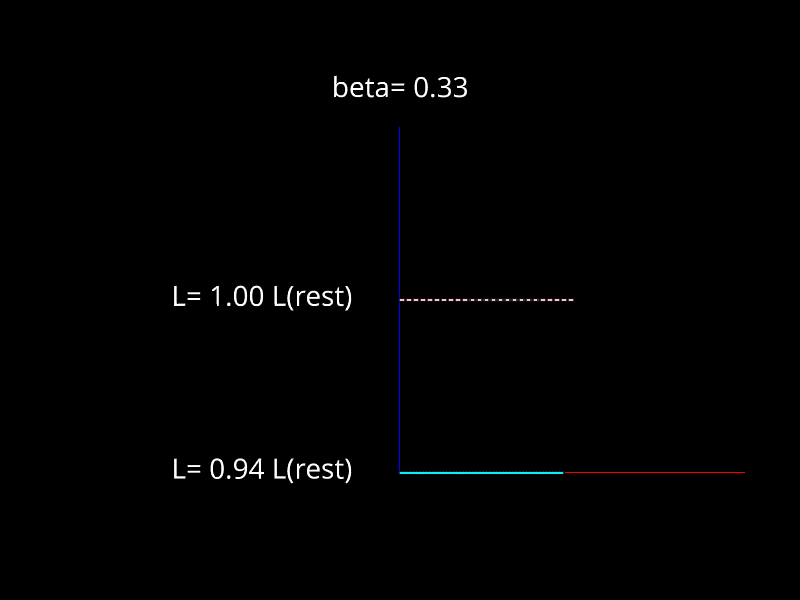}
	\label{fig:B1b}}
\quad
\subfloat[Subfigure 2 list of figures text][$\beta = 0.66$]{
	\includegraphics[width=0.24\textwidth]{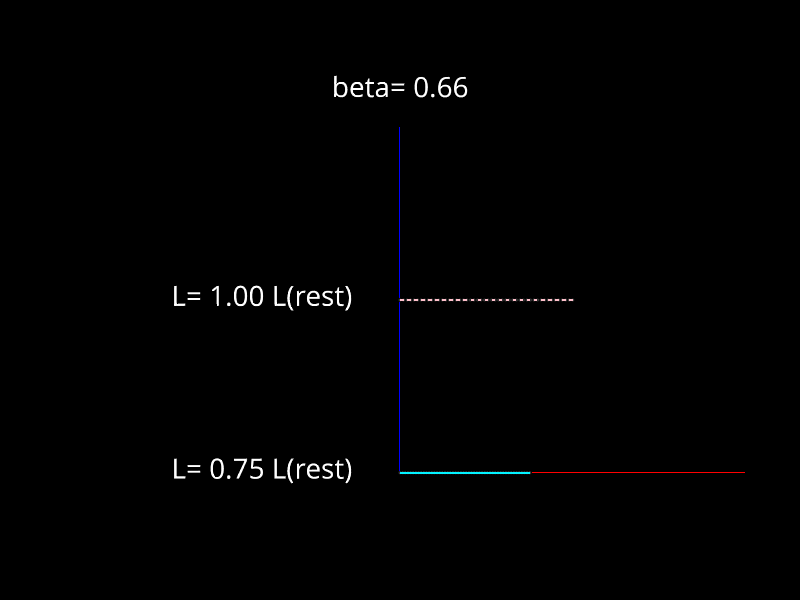}
	\label{fig:B1c}}
\subfloat[Subfigure 2 list of figures text][ $\beta = 0.99$ ]{
	\includegraphics[width=0.24\textwidth]{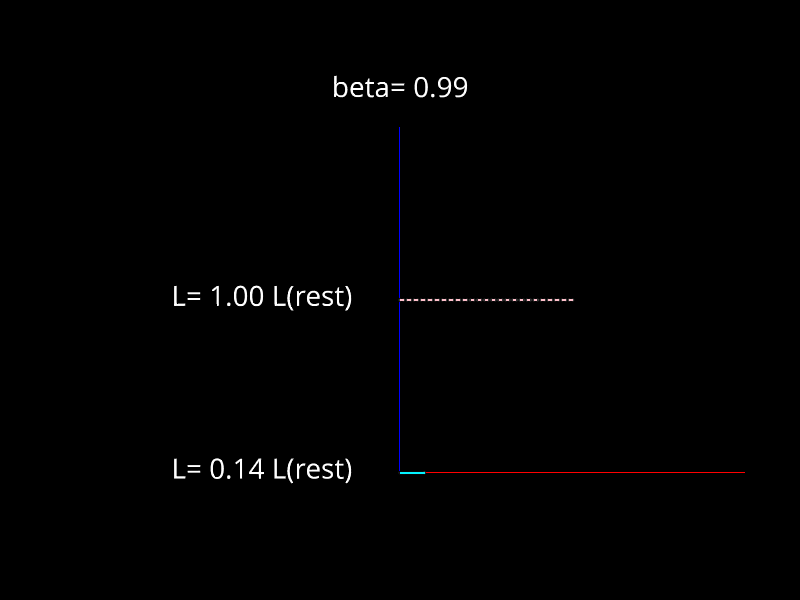}
	\label{fig:B1d}}
\caption{Lorentz transformed rod (cyan) oriented parallel to its motion compared to rod (dotted pink) at rest}
\end{figure}

\begin{figure}[H]
	\centering
	\subfloat[Subfigure 1 list of figures text][$\beta = 0.00$]{
		\includegraphics[width=0.24\textwidth]{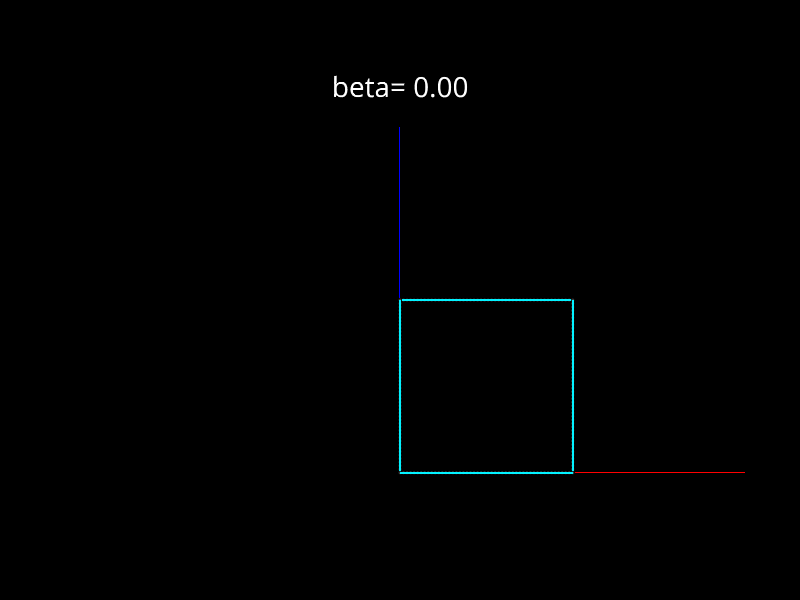}
		\label{fig:B2a}}
	\subfloat[Subfigure 2 list of figures text][$\beta = 0.33$]{
		\includegraphics[width=0.24\textwidth]{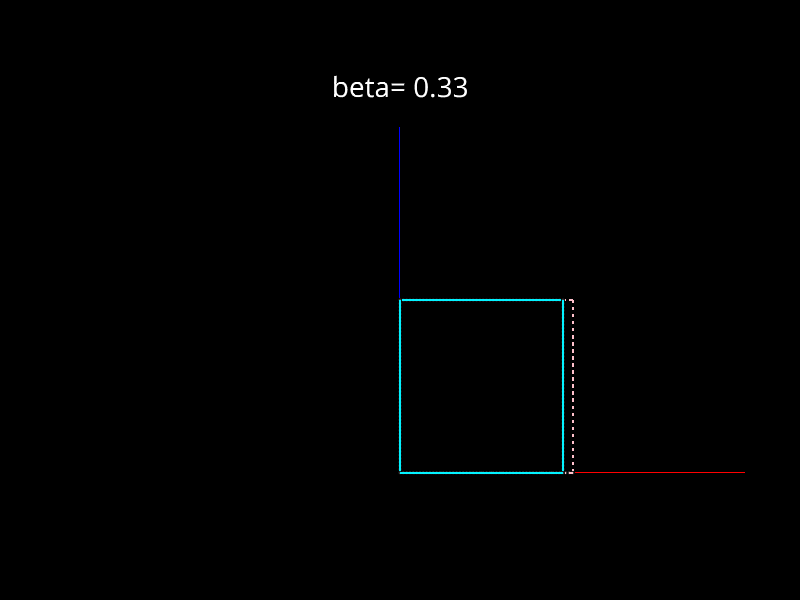}
		\label{fig:B2b}}
	\quad
	\subfloat[Subfigure 2 list of figures text][$\beta = 0.66$]{
		\includegraphics[width=0.24\textwidth]{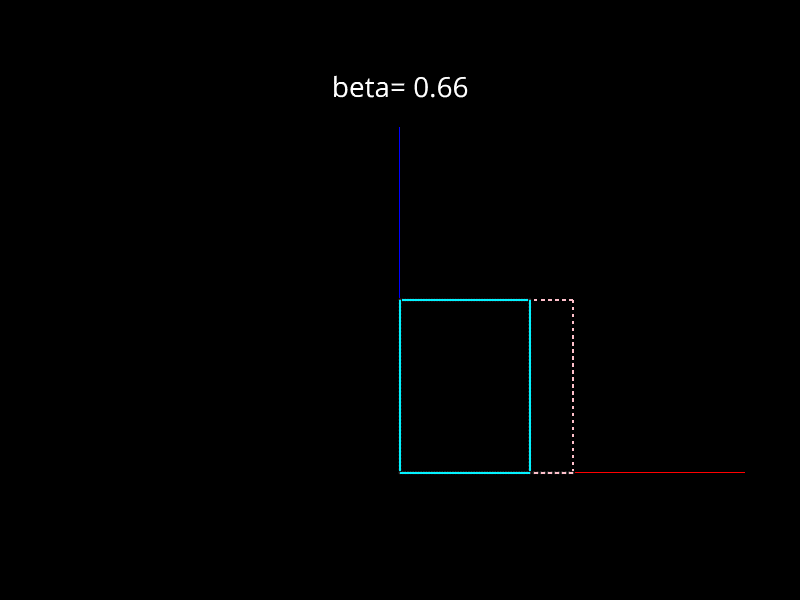}
		\label{fig:B2c}}
	\subfloat[Subfigure 2 list of figures text][$\beta = 0.99$]{
		\includegraphics[width=0.24\textwidth]{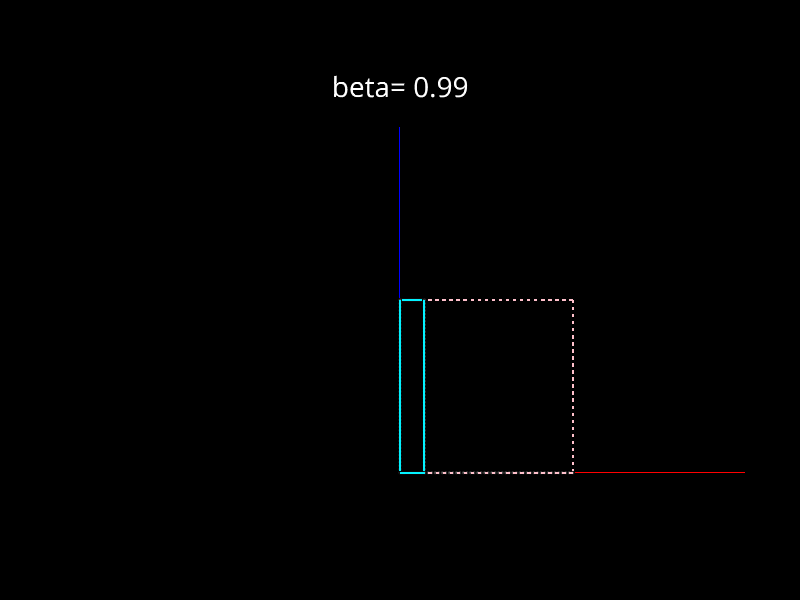}
		\label{fig:B2d}}
	\caption{Lorentz transformed square (cyan) oriented with $x$-direction parallel to motion compared to square at rest (dotted pink) }
\end{figure}
\begin{figure}[H]
	\vspace{13pt}
	\centering
	\subfloat[][$\beta = 0.00$]{
		\includegraphics[width=0.24\textwidth]{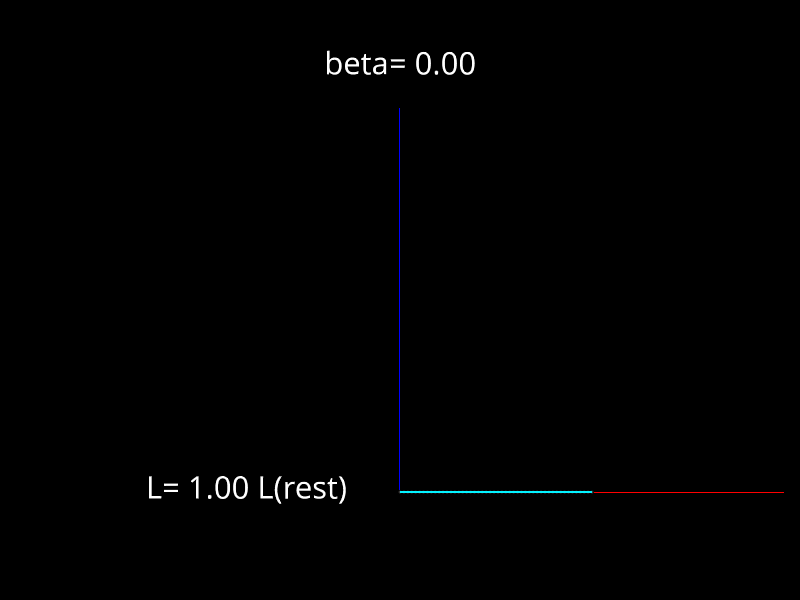}
		\label{fig:B3a}}
	\subfloat[][$\beta = 0.33$]{
		\includegraphics[width=0.24\textwidth]{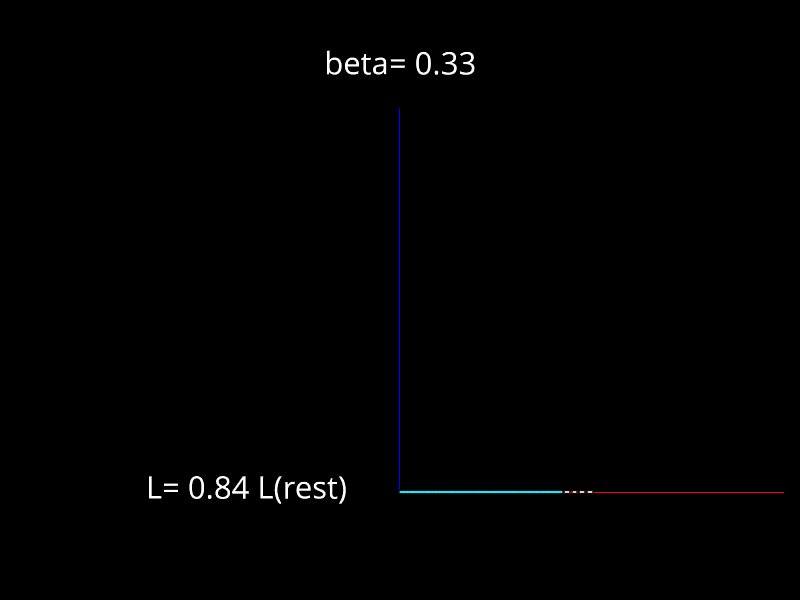}
		\label{fig:B3b}}
	\quad
	\subfloat[][$\beta = 0.66$]{
		\includegraphics[width=0.24\textwidth]{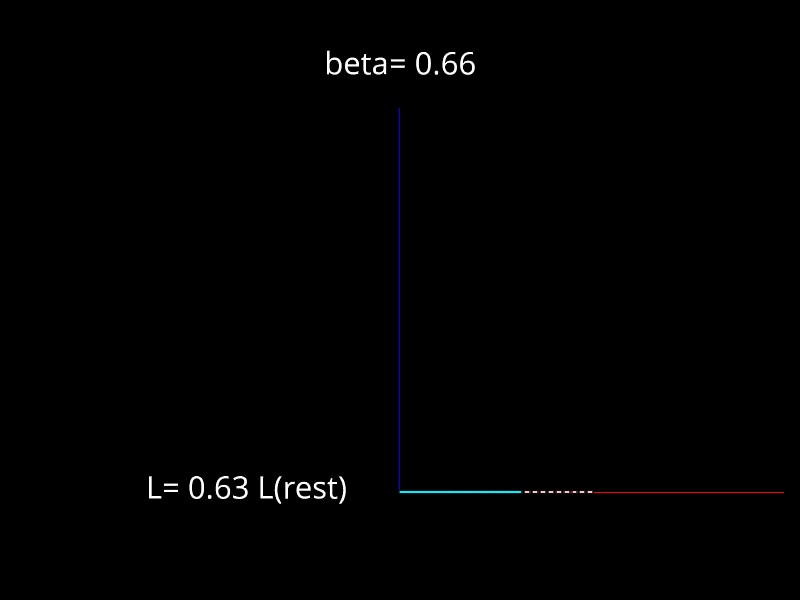}
		\label{fig:B3c}}
	\subfloat[][$\beta = 0.99$]{
		\includegraphics[width=0.24\textwidth]{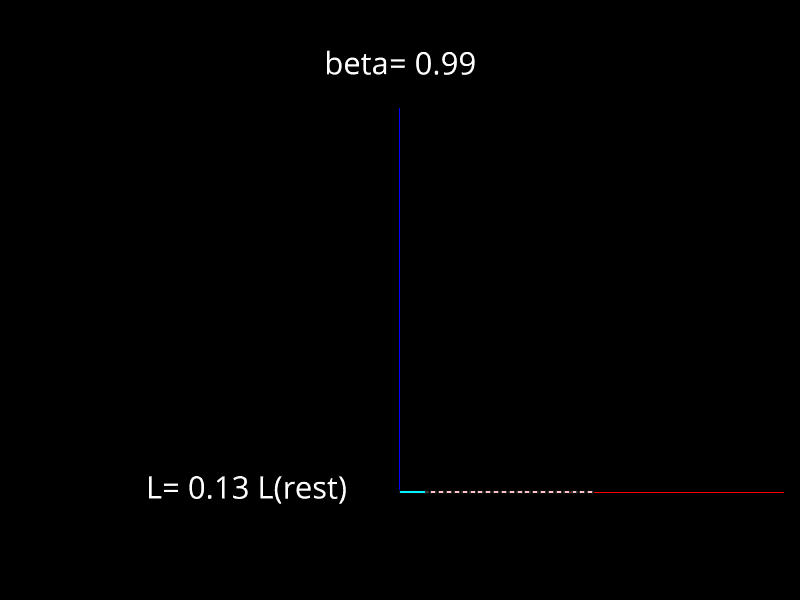}
		\label{fig:B3d}}
	\caption{Rod (cyan) oriented with $x$-direction parallel to motion (Lorentz and optical distortion) compared to rod at rest (dotted pink) }
\end{figure}

\begin{figure}[H]
	\centering
	\subfloat[][$\beta = 0.00$]{
		\includegraphics[width=0.24\textwidth]{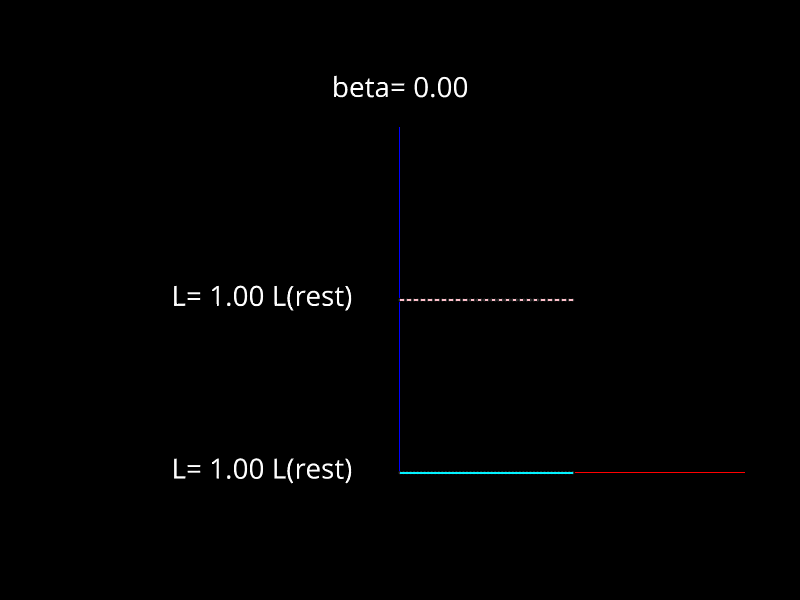}
		\label{fig:B4a}}
	\subfloat[][$\beta = 0.33$]{
		\includegraphics[width=0.24\textwidth]{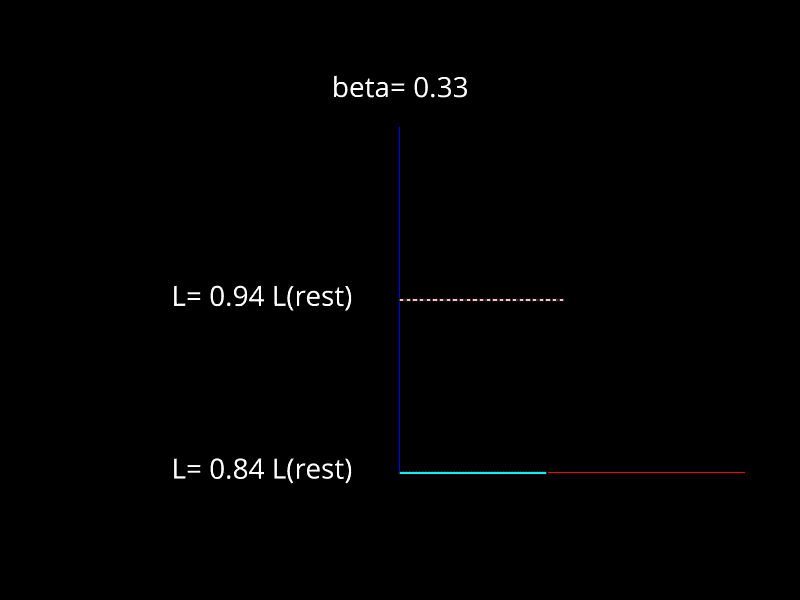}
		\label{fig:B4b}}
	\quad
	\subfloat[][$\beta = 0.66$]{
		\includegraphics[width=0.24\textwidth]{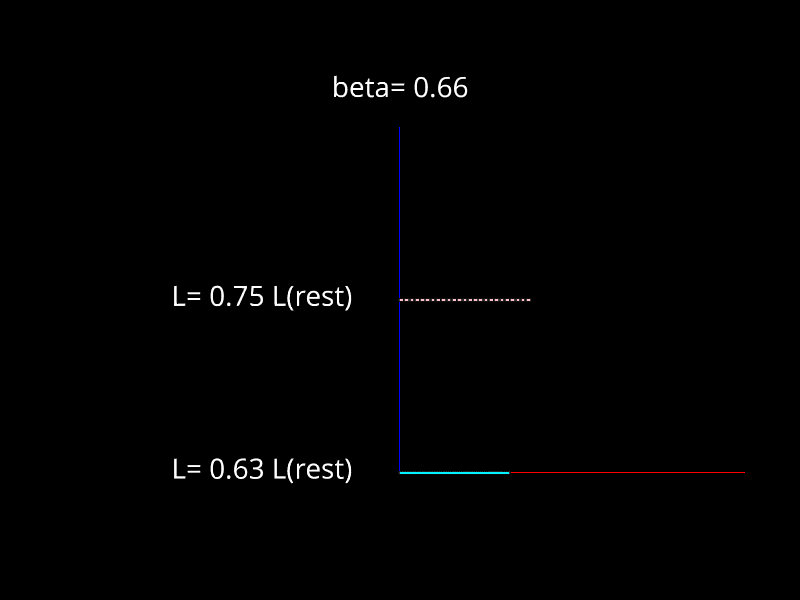}
		\label{fig:B4c}}
	\subfloat[][$\beta = 0.99$]{
		\includegraphics[width=0.24\textwidth]{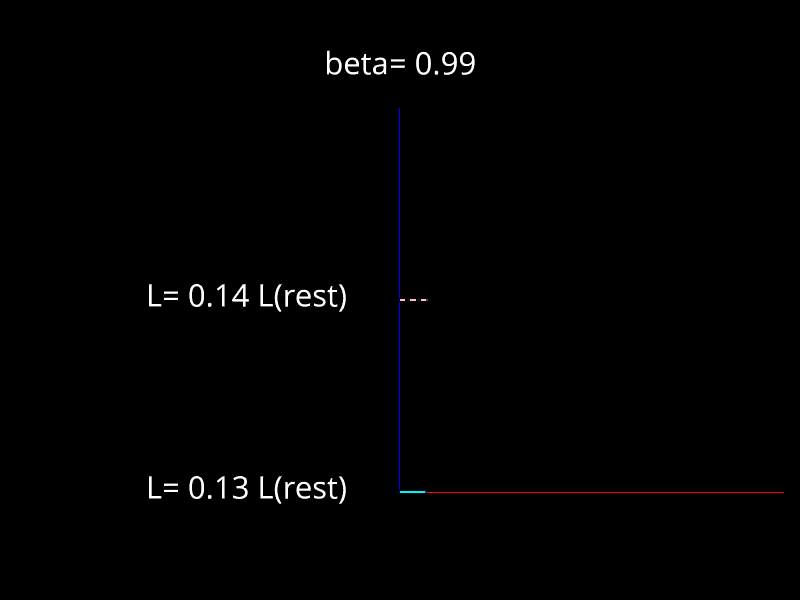}
		\label{fig:B4d}}
	\caption{Comparison of Lorentz contraction (pink) to optical distortion (cyan) in rods oriented parallel to motion }
\end{figure}
\afterpage{\clearpage}
\begin{figure}[H]
	\centering
	\subfloat[][$\beta = 0.00$]{
		\includegraphics[width=0.24\textwidth]{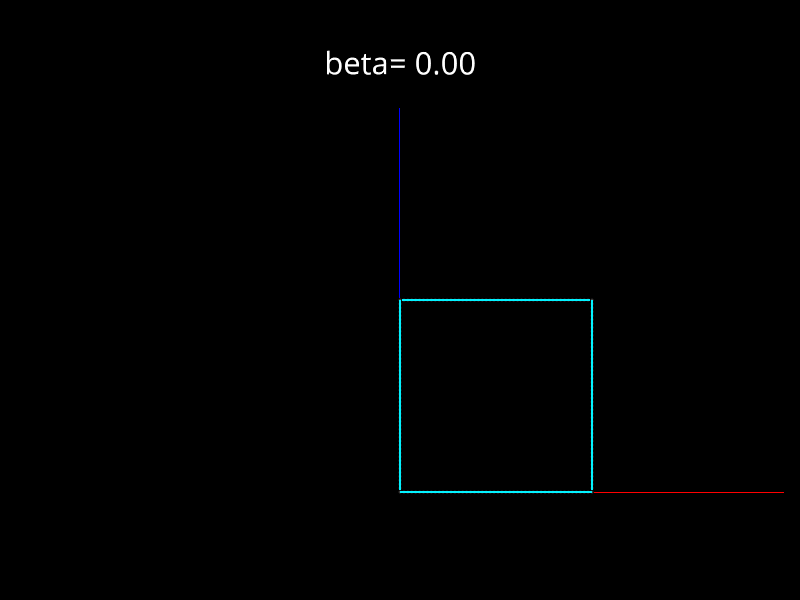}
		\label{fig:B5a}}
	\subfloat[][$\beta = 0.33$]{
		\includegraphics[width=0.24\textwidth]{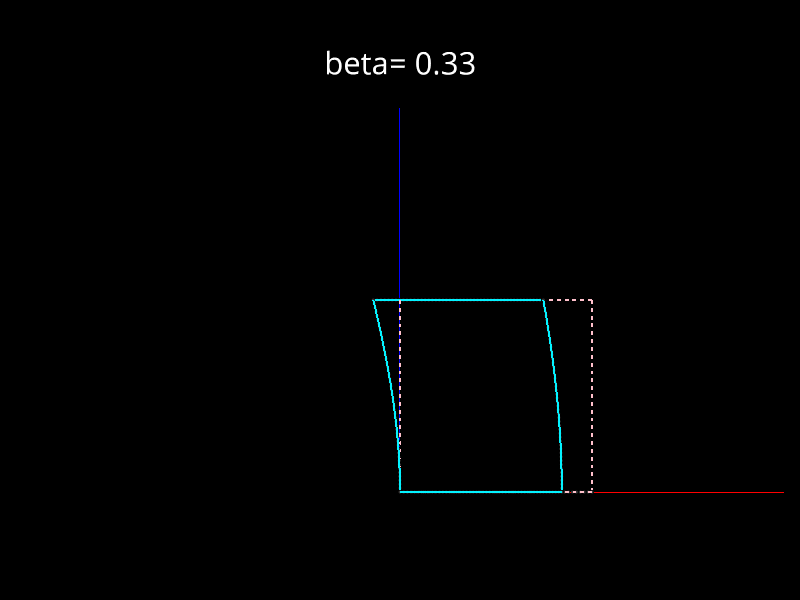}
		\label{fig:B5b}}
	\quad
	\subfloat[][$\beta = 0.66$]{
		\includegraphics[width=0.24\textwidth]{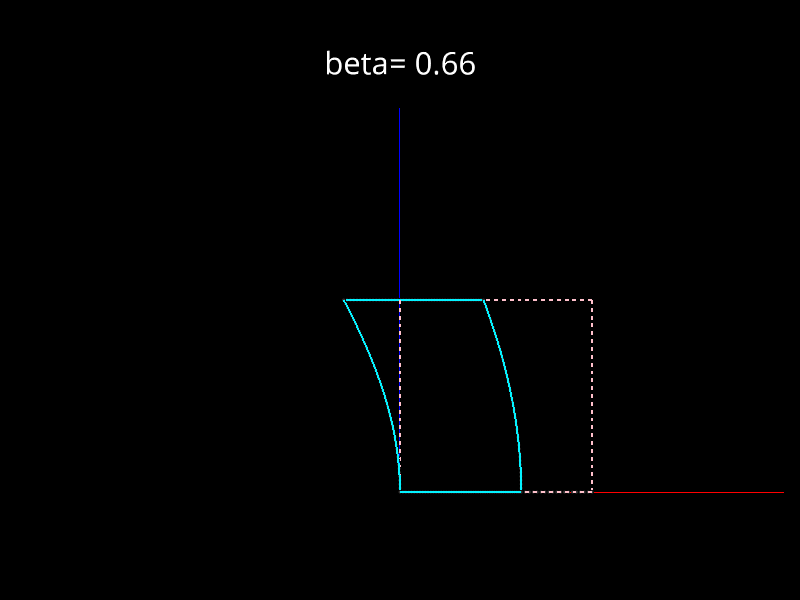}
		\label{fig:B5c}}
	\subfloat[][$\beta = 0.99$]{
		\includegraphics[width=0.24\textwidth]{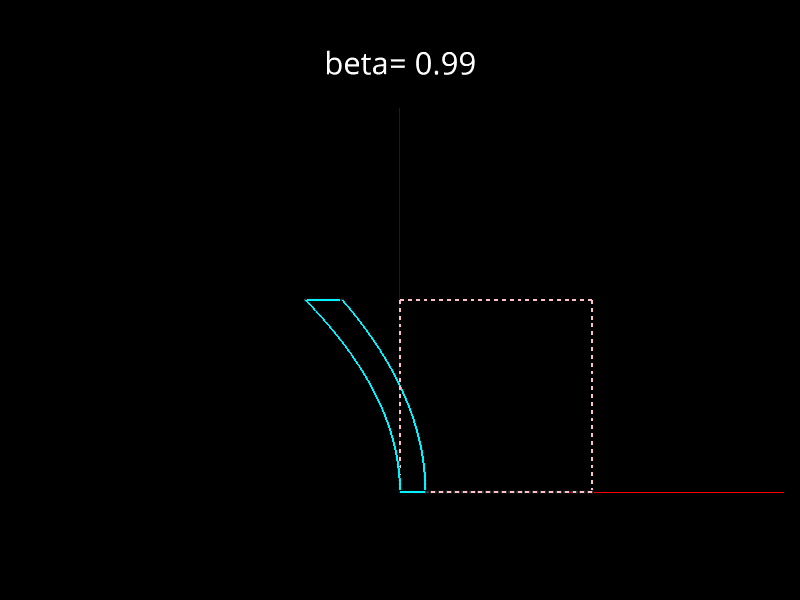}
		\label{fig:B5d}}
	\caption{Square (cyan) oriented with $x$-direction parallel to motion (Lorentz and optical distortion) compared to square at rest (dotted pink) }
\end{figure}
\begin{figure}[H]
	\centering
	\subfloat[][$\beta = 0.00$]{
		\includegraphics[width=0.24\textwidth]{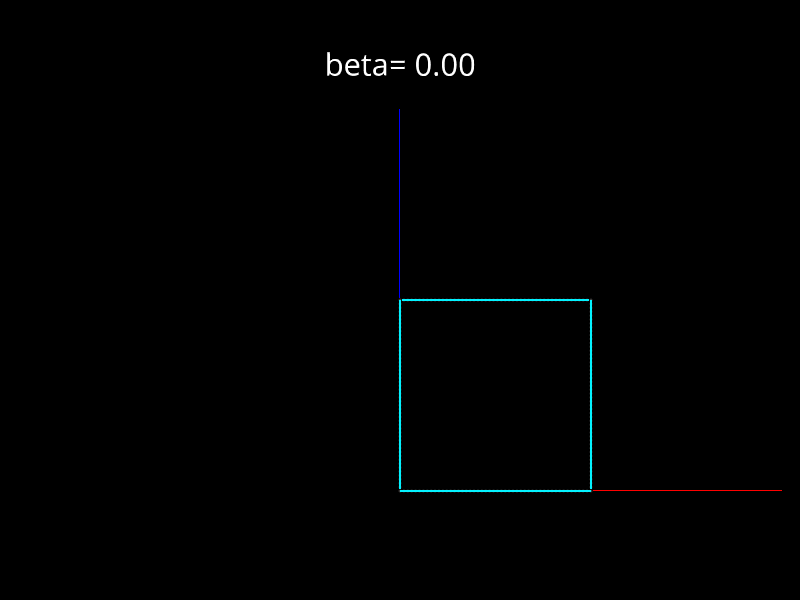}
		\label{fig:B6a}}
	\subfloat[][$\beta = 0.33$]{
		\includegraphics[width=0.24\textwidth]{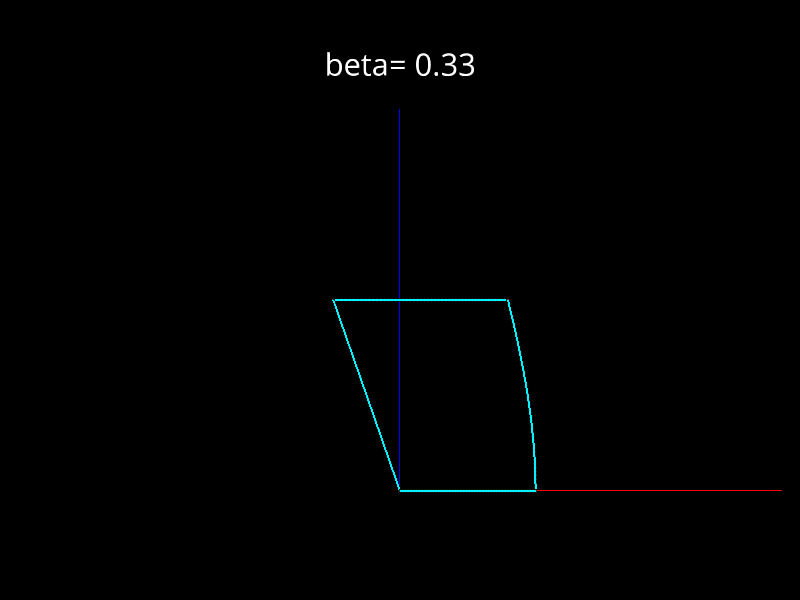}
		\label{fig:B6b}}
	\quad
	\subfloat[][$\beta = 0.66$]{
		\includegraphics[width=0.24\textwidth]{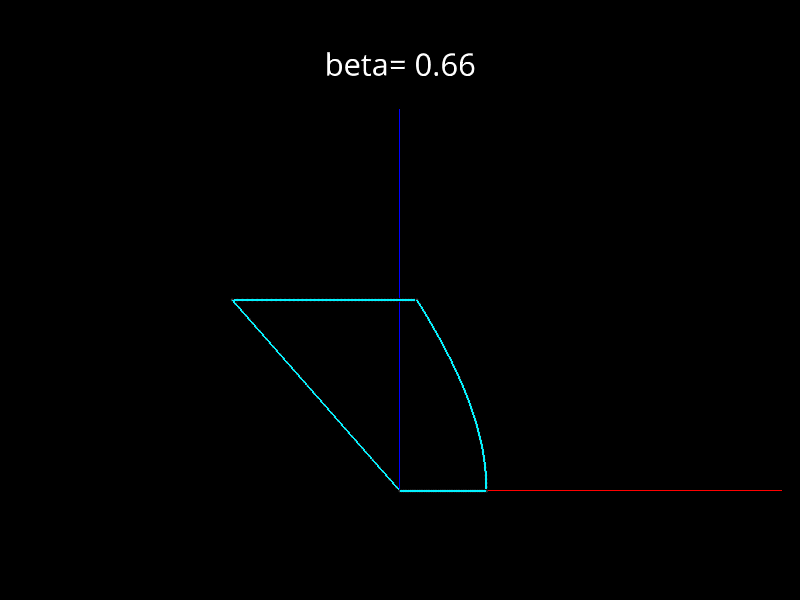}
		\label{fig:B6c}}
	\subfloat[][$\beta = 0.99$]{
		\includegraphics[width=0.24\textwidth]{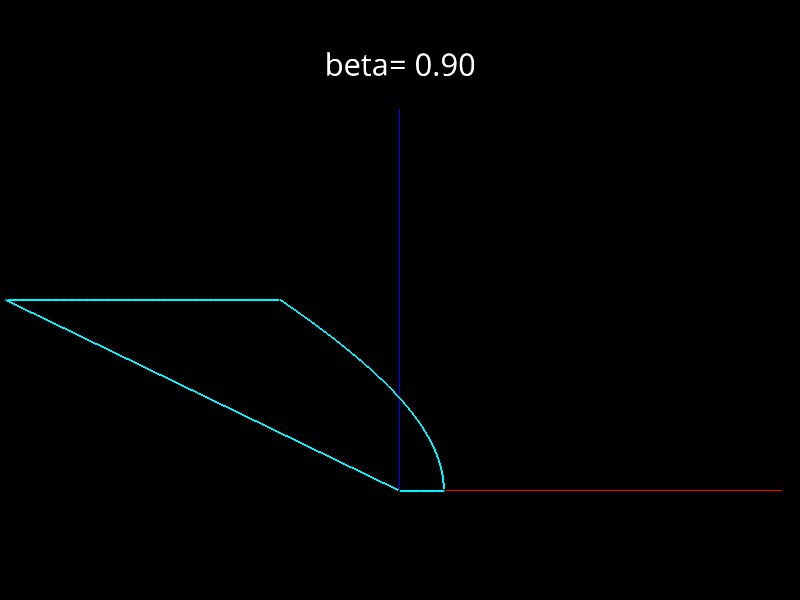}
		\label{fig:B6d}}
	\caption{Degenerate case of a square viewed with camera position $d_{c} = (0, 0, 0)$}
\end{figure}
\begin{figure}[H]
	\centering
	\subfloat[][$\beta = 0.00$]{
		\includegraphics[width=0.24\textwidth]{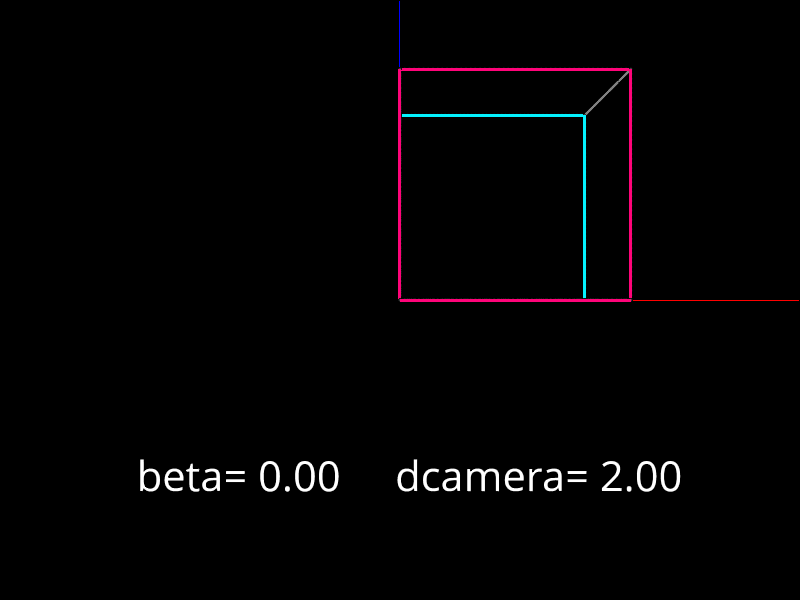}
		\label{fig:B7a}}
	\subfloat[][$\beta = 0.33$]{
		\includegraphics[width=0.24\textwidth]{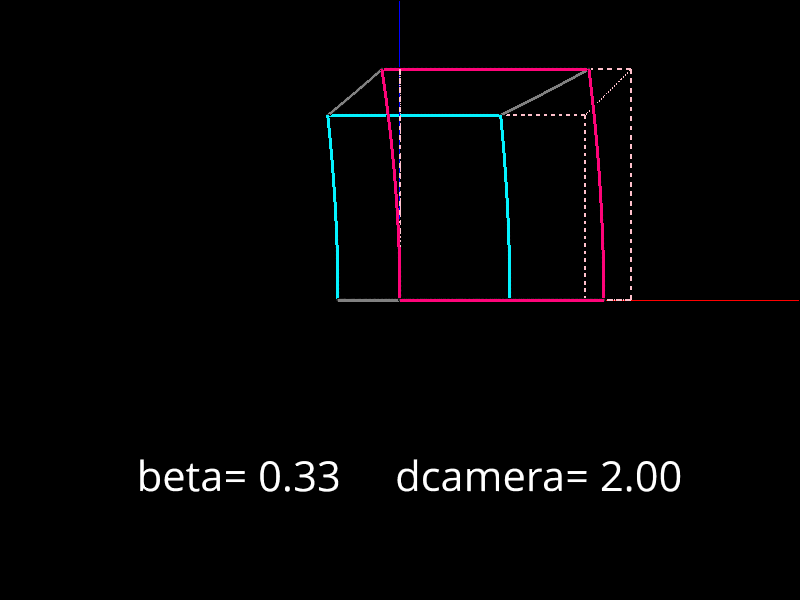}
		\label{fig:B7b}}
	\quad
	\subfloat[][$\beta = 0.66$]{
		\includegraphics[width=0.24\textwidth]{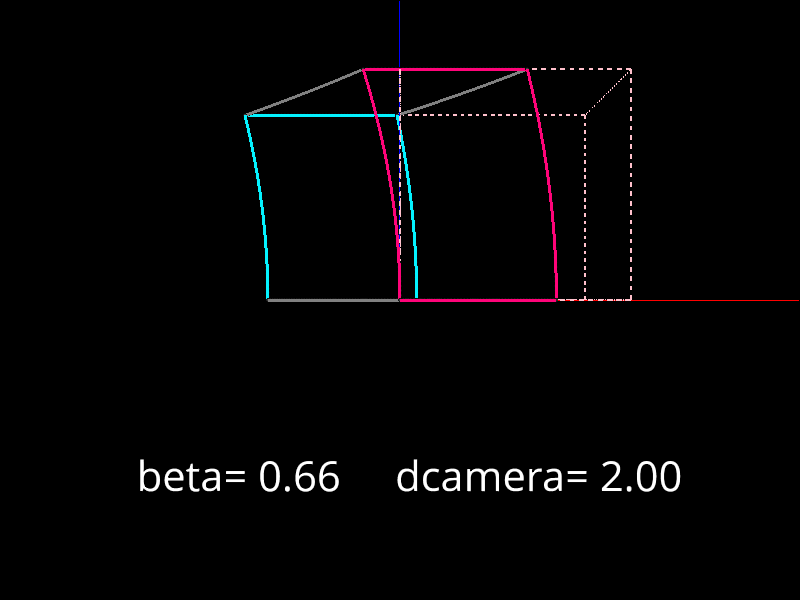}
		\label{fig:B7c}}
	\subfloat[][$\beta = 0.99$]{
		\includegraphics[width=0.24\textwidth]{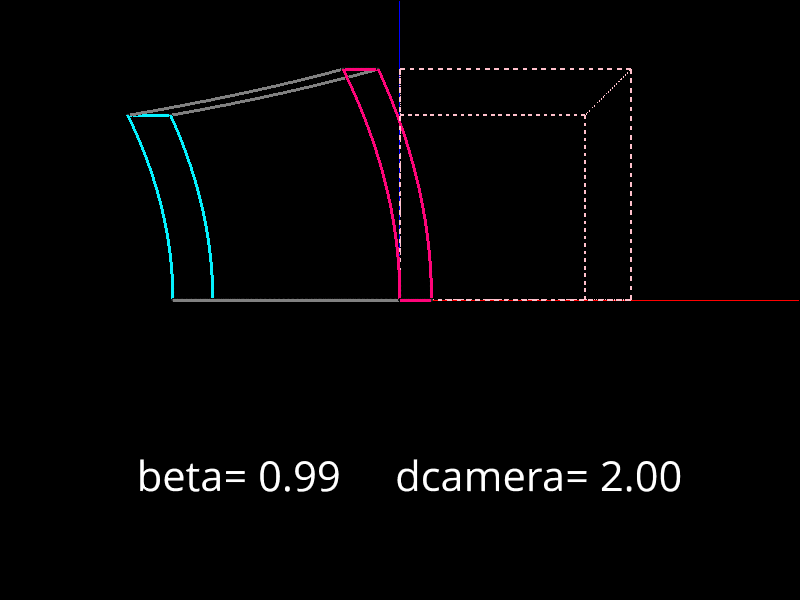}
		\label{fig:B7d}}
	\caption{cube oriented with $x$-direction parallel to motion (Penrose-Terrel Rotation) compared to cube at rest (dotted pink) }
\end{figure}
\begin{figure}[H]
	\centering
	\subfloat[][$\beta = 0.00$]{
		\includegraphics[width=0.24\textwidth]{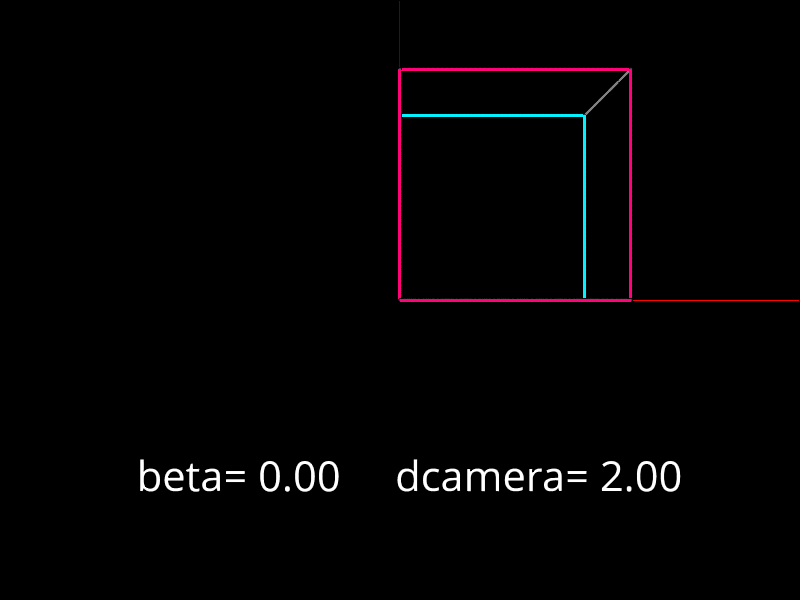}
		\label{fig:B8a}}
	\subfloat[][$\beta = 0.33$]{
		\includegraphics[width=0.24\textwidth]{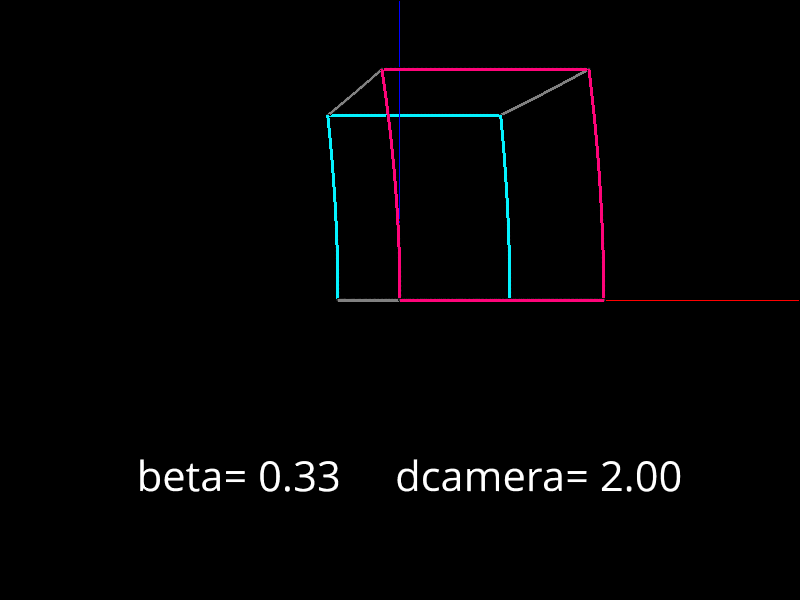}
		\label{fig:B8b}}
	\quad
	\subfloat[][$\beta = 0.66$]{
		\includegraphics[width=0.24\textwidth]{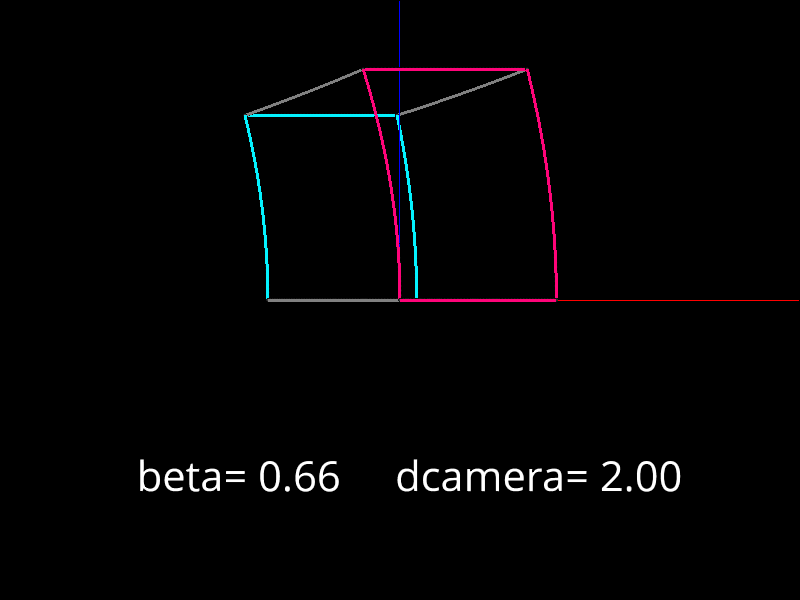}
		\label{fig:B8c}}
	\subfloat[][$\beta = 0.99$]{
		\includegraphics[width=0.24\textwidth]{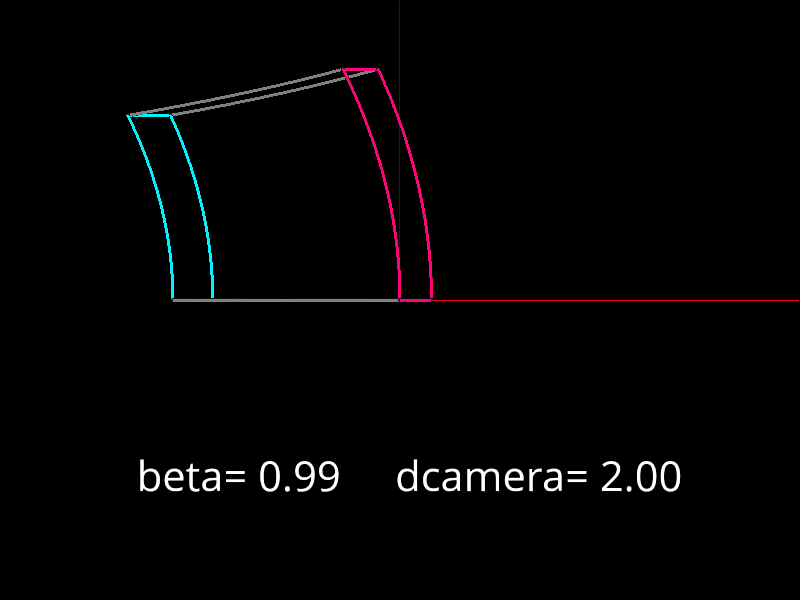}
		\label{fig:B8d}}
	\caption{Cube oriented with $x$-direction parallel to motion (Penrose-Terrel Rotation) without comparison }
\end{figure}
\afterpage{\clearpage}
\begin{figure}[H]
	\centering
	\subfloat[][$\beta = 0.00$]{
		\includegraphics[width=0.24\textwidth]{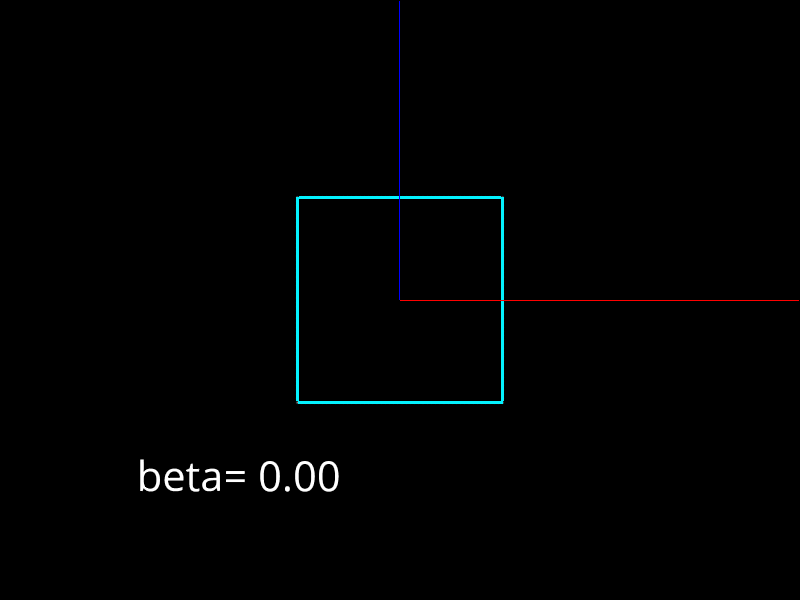}
		\label{fig:B9a}}
	\subfloat[][$\beta = 0.33$]{
		\includegraphics[width=0.24\textwidth]{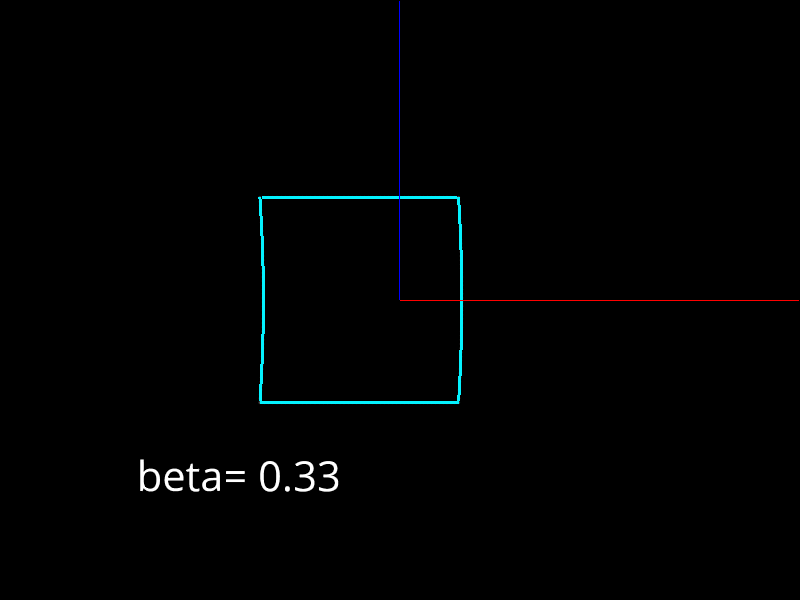}
		\label{fig:B9b}}
	\quad
	\subfloat[][$\beta = 0.66$]{
		\includegraphics[width=0.24\textwidth]{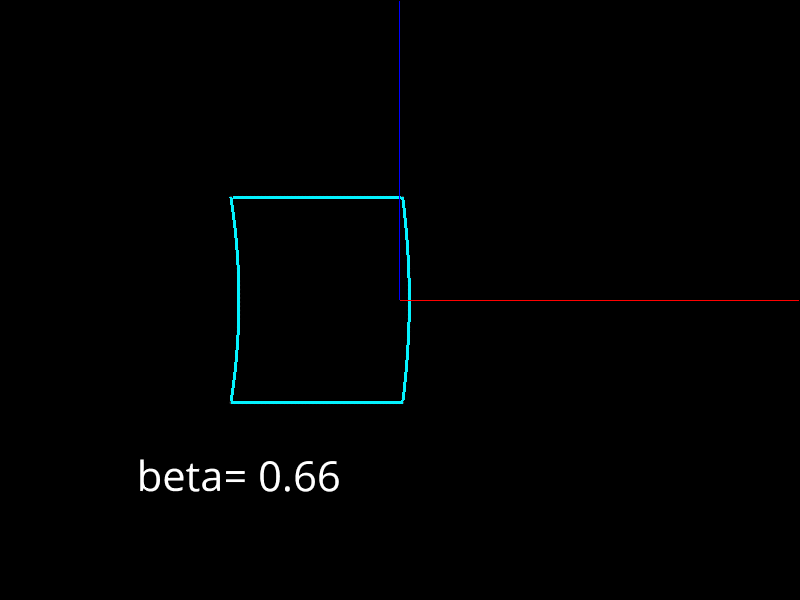}
		\label{fig:B9c}}
	\subfloat[][$\beta = 0.99$]{
		\includegraphics[width=0.24\textwidth]{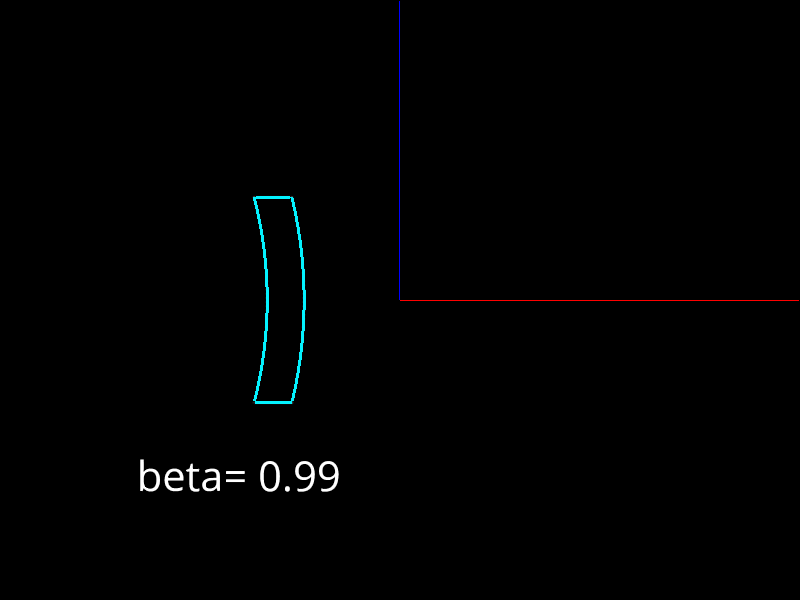}
		\label{fig:B9d}}
	\caption{Square centered at origin with Optical Distortion}
\end{figure}
\begin{figure}[H]
	\centering
	\subfloat[][$\beta = 0.00$]{
		\includegraphics[width=0.24\textwidth]{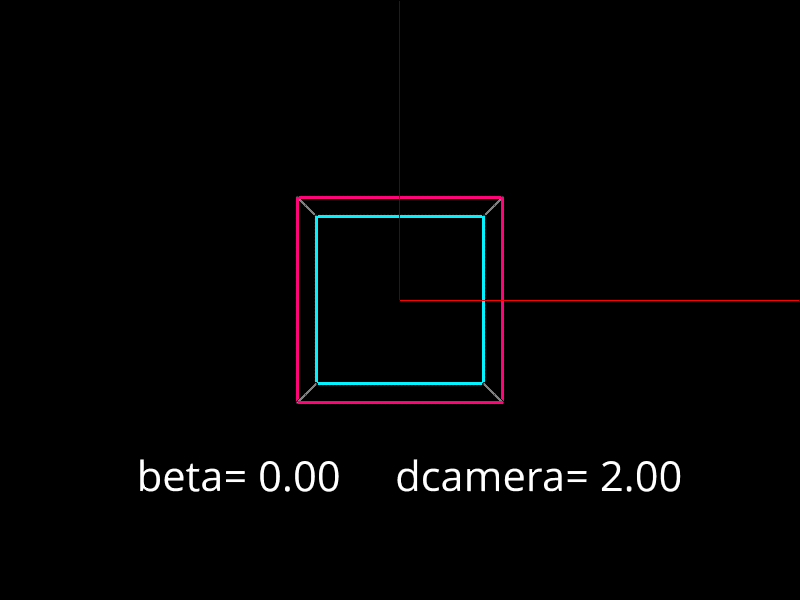}
		\label{fig:B10a}}
	\subfloat[][$\beta = 0.33$]{
		\includegraphics[width=0.24\textwidth]{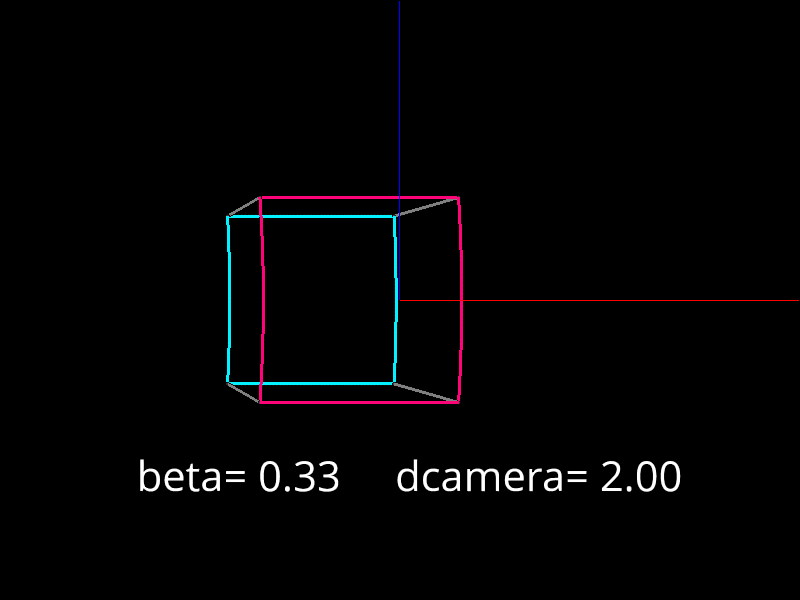}
		\label{fig:B10b}}
	\quad
	\subfloat[][$\beta = 0.66$]{
		\includegraphics[width=0.24\textwidth]{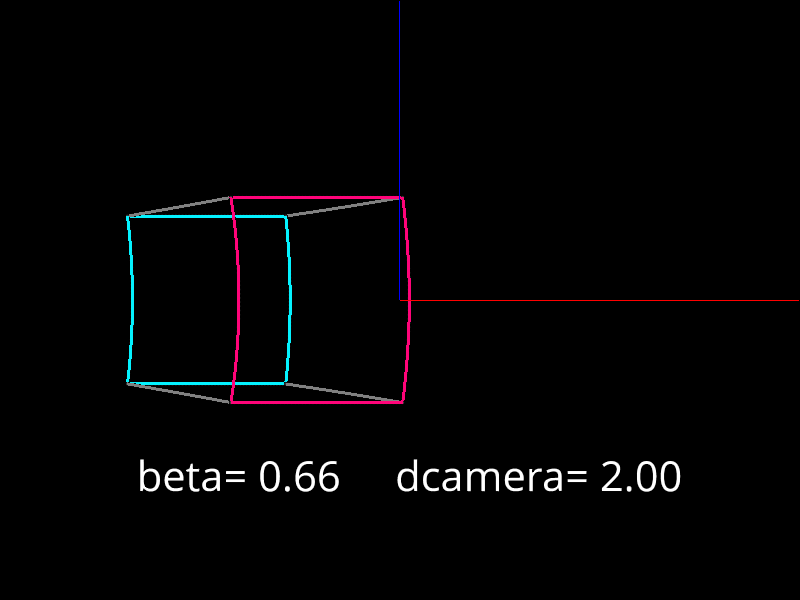}
		\label{fig:B10c}}
	\subfloat[][$\beta = 0.99$]{
		\includegraphics[width=0.24\textwidth]{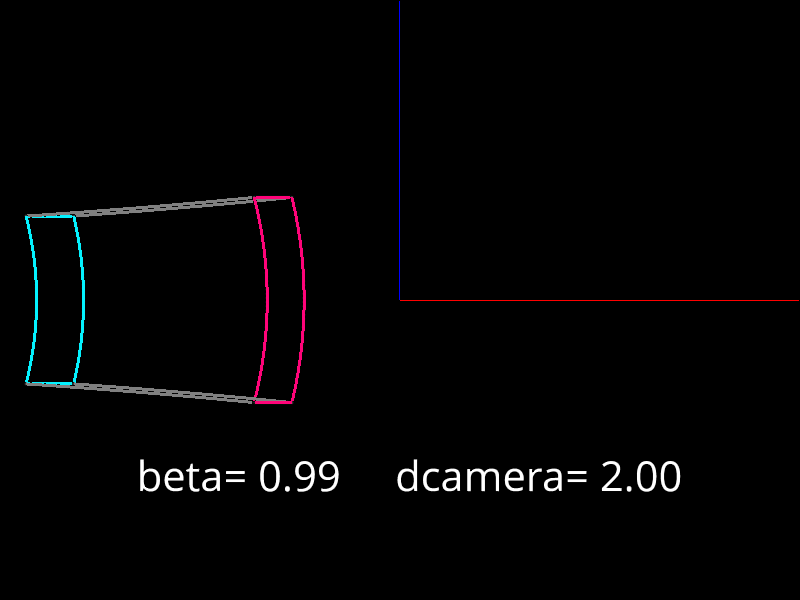}
		\label{fig:B10d}}
	\caption{Cube centered at origin with Penrose-Terrell Rotation}
\end{figure}
\begin{widetext}
	\begin{figure*}[b]
		\centering
	\subfloat[][$\beta = 0.66$]{
		\includegraphics[width=0.33\textwidth]{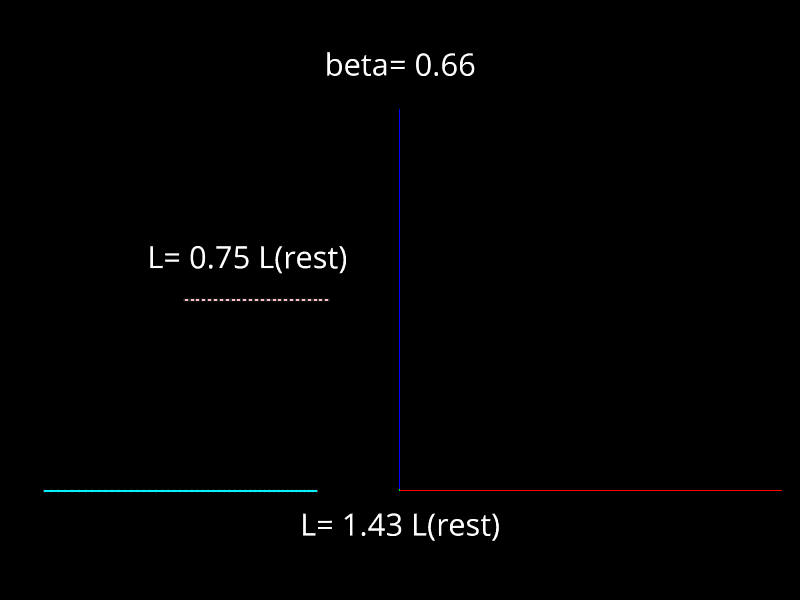}
			\label{fig:B11a}}
	\subfloat[][$\beta = 0.66$]{
		\includegraphics[width=0.33\textwidth]{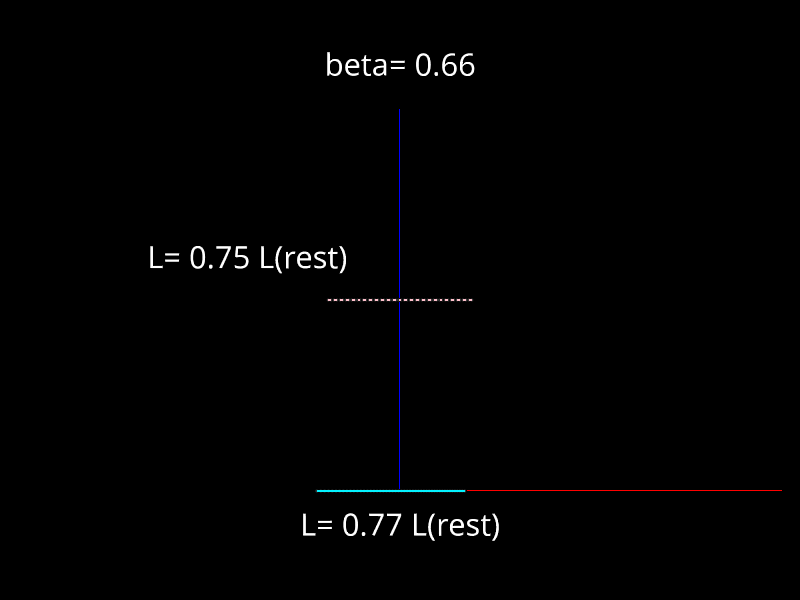}
			\label{fig:B11b}}
	\subfloat[][$\beta = 0.66$]{
		\includegraphics[width=0.33\textwidth]{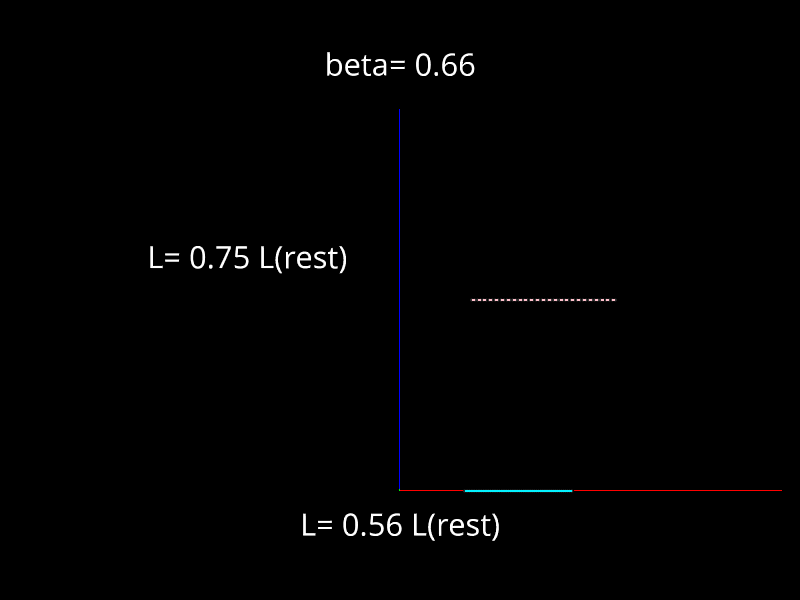}
			\label{fig:B11c}}
		\quad
	\subfloat[][$\beta = 0$]{
		\includegraphics[width=0.33\textwidth]{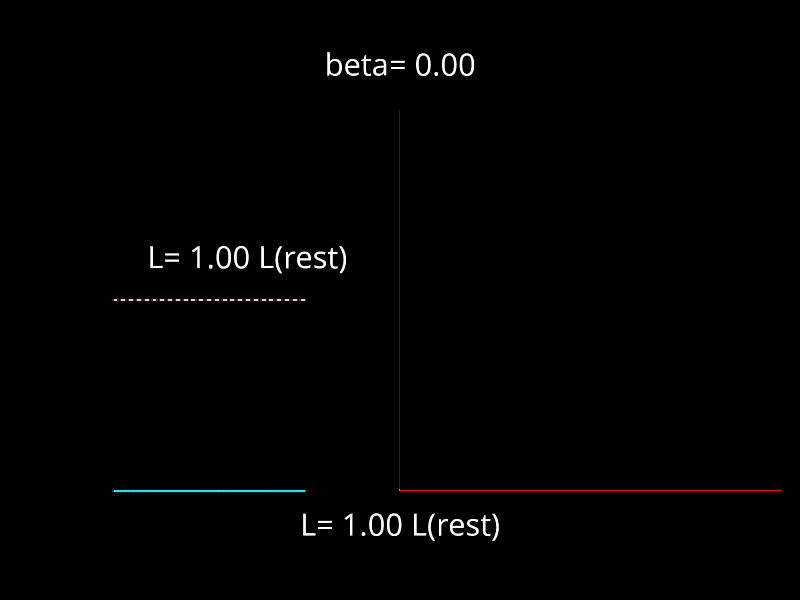}
		\label{fig:B11d}}
	\subfloat[][$\beta = 0$]{
		\includegraphics[width=0.33\textwidth]{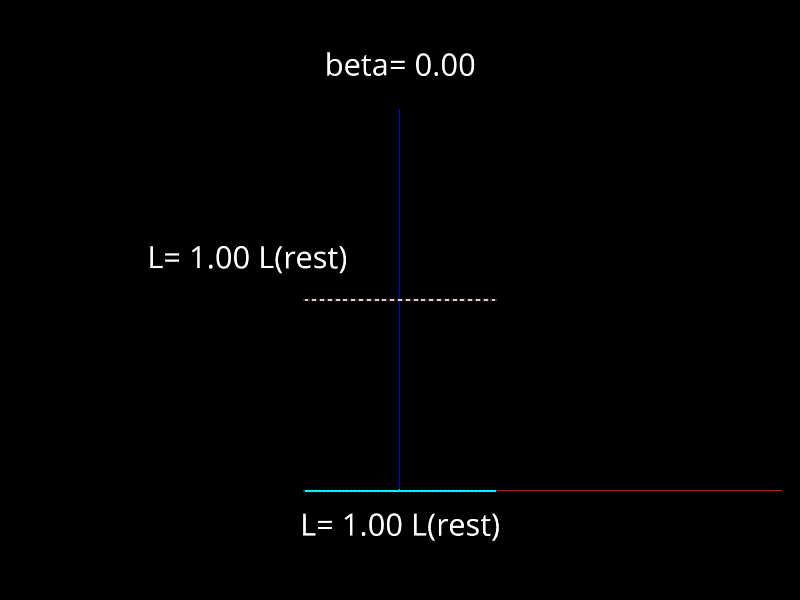}
		\label{fig:B11e}}
	\subfloat[][$\beta = 0$]{
		\includegraphics[width=0.33\textwidth]{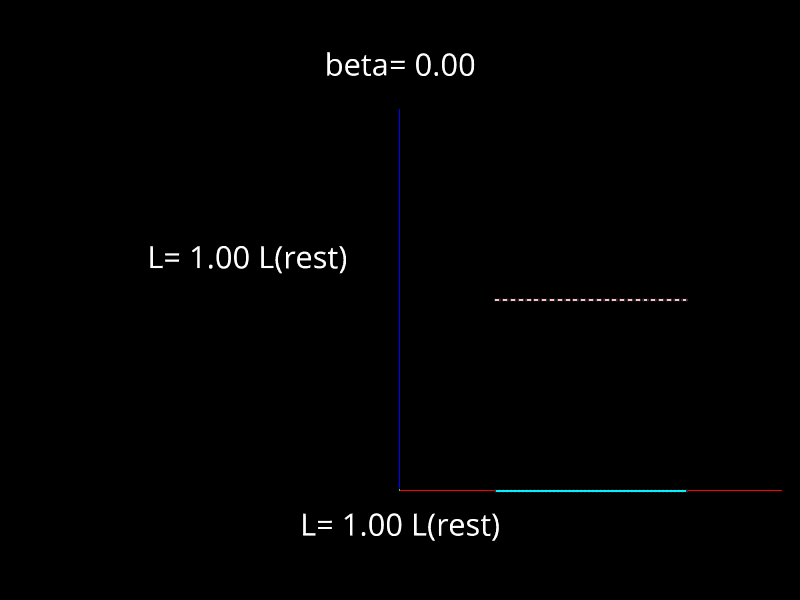}
		\label{fig:B11f}}
		\caption{Spatial dependence on optical distortion effects (cyan) compared to Loretnz contraction (pink dotted). Due to the geometry, the optically distorted rod may appear to be longer than its rest length when approaching the observer and is always longer than the Lorentz contracted length when oriented to the left.}
	\end{figure*}
\newpage
\end{widetext}
\bibliography{thesisbib}

\end{document}